\documentclass[11pt]{article}
\usepackage{epsfig}

\def\text{}
\renewcommand{\baselinestretch}{1.75} 

\newcommand{\be}{\begin{eqnarray}}
\newcommand{\ee}{\end{eqnarray}}

\newcommand{\np}{\newpage}

\newcommand{\vs}{\vspace}

\newcommand{\nn}{\nonumber}

\def\so#1{{\rm SO}( #1)}
\def\su#1{{\rm SU}( #1)}
\def\sp#1{{\rm Sp}( #1)}
\def\u#1{{\rm U}( #1)}
\def\ret{\nonumber\\{}}
\def\la#1{\label{#1}}
\def\eqs#1{(\ref{#1})}

\def\half{\textstyle\frac{1}{2}} 
\def\C#1{{\cal #1}}
\def\rep#1{{\bf #1}}
\font\upright=cmu10 
\font\cmss=cmss10 at 11pt \font\cmsss=cmss8 at 8pt

\font\mytinyssf=cmss10 at 14pt

\def\inbar{\vrule height1.5ex width.4pt depth0pt}
\def\mininbar{\vrule height.75ex width.3pt depth0pt} 
\def\cc{\relax\,\hbox{$\mininbar\kern-.2em{\hbox{\rm\tiny
C}}$}}
\def\IZ{\relax\ifmmode\mathchoice
{\hbox{\cmss Z\kern-.4em Z}}{\hbox{\cmss Z\kern-.4em Z}}
{\lower.4pt\hbox{\cmsss Z\kern-.4em Z}}
{\lower1.2pt\hbox{\cmsss Z\kern-.4em Z}}\else{\cmss Z\kern-.4em
Z}\fi}
\def\IC{\relax\,\hbox{$\inbar\kern-.3em{\rm C}$}}
\def\IR{\relax{\rm I\kern-.18em R}}
\def\mt{\rlap{\cmss T}\kern 3.0pt{\hbox{{\cmss T}}}}
\def\sss#1{\hbox{\mytinyssf #1}}
\def\identity{{\upright\rlap{1}\kern 2.0pt 1}}
\def\globalmass{{\sf m}}
\def\mod{~{\rm mod}~}
\def\sqr#1#2{{\vcenter{\vbox{\hrule height.#2pt
    \hbox{\vrule width.#2pt height#1pt \kern#1pt
    \vrule width.#2pt}
    \hrule height.#2pt}}}}

\newdimen\tableauside\tableauside=1.0ex
\newdimen\tableaurule\tableaurule=0.4pt
\newdimen\tableaustep
\def\phantomhrule#1{\hbox{\vbox to0pt{\hrule height\tableaurule
width#1\vss}}}
\def\phantomvrule#1{\vbox{\hbox to0pt{\vrule width\tableaurule
height#1\hss}}}
\def\sqr{\vbox{%
  \phantomhrule\tableaustep
 
\hbox{\phantomvrule\tableaustep\kern\tableaustep\phantomvrule\tableaustep}%
  \hbox{\vbox{\phantomhrule\tableauside}\kern-\tableaurule}}}
\def\squares#1{\hbox{\count0=#1\noindent\loop\sqr
  \advance\count0 by-1 \ifnum\count0>0\repeat}}
\def\tableau#1{\vcenter{\offinterlineskip
  \tableaustep=\tableauside\advance\tableaustep by-\tableaurule
  \kern\normallineskip\hbox
    {\kern\normallineskip\vbox
      {\gettableau#1 0 }%
     \kern\normallineskip\kern\tableaurule}%
  \kern\normallineskip\kern\tableaurule}}
\def\gettableau#1 {\ifnum#1=0\let\next=\null\else
  \squares{#1}\let\next=\gettableau\fi\next}
\newcommand{\fund}{\tableau{1}}
\newcommand{\Ysymm}{\tableau{2}}
\newcommand{\Yasymm}{\tableau{1 1}}
\def\tablecomp{1} 
\def\tablescalars{2}
\def\tablevectors{3}
\def\tblspso{4}
 \def\tblbranching{5}
\def\tblspanti{6}
\def\tblspantifield{7}
\def\tbltensor{8}
\def\tblsutwoanti{9}
\def\tblsutwoantifield{10}
\def\tblsutwoantivect{11}
 \def\tblspxsp{12}
 \def\tblspxspfield{13}
 \def\tblsuantisym{14}
 \def\tblsuantisymfield{15}
 \def\tblspxso{16}
 \def\tblspxsofield{17}

\begin{document}

\thispagestyle{empty}

\vs*{-25mm}
\begin{flushright}
BRX-TH-474\\[-.15in]
BOW-PH-119\\[-.15in] 
HUTP-00/A019\\[-.15in]
\end{flushright}
\vspace{.3in}
\setcounter{footnote}{0}

\begin{center}
{\huge{\bf Elliptic models, type IIB orientifolds, \\
and the AdS/CFT
correspondence}}
\renewcommand{\baselinestretch}{1}
\small
\normalsize
\vspace{.5in}

Isabel P. Ennes\footnote{
Research supported 
by the DOE under grant DE--FG02--92ER40706.}$^{,a}$, 
Carlos Lozano\footnotemark[1]$^{,a}$, Stephen G. Naculich\footnote{
Research supported in part by the National Science Foundation under grant 
no.~PHY94-07194 through the \\
\phantom{aaa}  ITP Scholars Program.}$^{,b}$, 
Howard J. Schnitzer\footnote{Permanent address.}
${}^{\!\!\!,\!\!\!}$
\footnote{Research supported in part
by the DOE under grant DE--FG02--92ER40706.\\
{\tt \phantom{aaa} naculich@bowdoin.edu; 
ennes,lozano,schnitzer@brandeis.edu}\\}$^{,a,c}$\\

\vspace{.2in}

$^{3,a}$Martin Fisher School of Physics\\
Brandeis University, Waltham, MA 02454

\vspace{.2in}

$^{b}$Department of Physics\\
Bowdoin College, Brunswick, ME 04011

\vspace{.2in}

$^{c}$Lyman Laboratory of Physics \\
Harvard University, Cambridge, MA 02138

\vspace{.3in}

{\bf{Abstract}} 
\end{center}
\renewcommand{\baselinestretch}{1.75}
\small
\normalsize
\begin{quotation}
\baselineskip14pt
\noindent  

We analyze the large $N$ supergravity descriptions 
of the class of type IIB models T-dual 
to elliptic type IIA brane configurations 
containing two orientifold 6-planes and up to two NS 5-branes.
The T-dual IIB configurations contain
$N$ D3-branes in the background of an orientifold 7-plane and, 
in some models, a $\IZ_2$ orbifold and/or D7-branes,
which give rise to four-dimensional $\C{N}=2$ (or $\C{N}=4$) gauge theories 
with at most two factors.   
We identify the chiral primary states of the supergravity theories,
and match them to gauge invariant operators 
of the corresponding superconformal theories using Maldacena's duality.
\end{quotation}

\np 

\setcounter{page}{1}

\baselineskip20pt
\setcounter{footnote}{0}

\tableofcontents

\section{Introduction}

Maldacena's duality conjecture 
\cite{MaldacenaAds}--\cite{AdsReview}  
relating string theories on AdS space
to conformal field theories on its boundary
has proved to be extremely fruitful.
The original conjecture involved the $\C{N}=4$
supersymmetric $\su{N}$ gauge theory 
on the worldvolume of $N$ D3-branes.
In the large $N$ limit, the near-horizon geometry of the
D3-branes is $AdS_5 \times S^5$. 
Additional structures may be added to the background,
leading to generalizations of the original proposal,
where the worldvolume theory of D3-branes probing
orbifold/orientifold backgrounds may be derived systematically
using string-theoretic methods \cite{GimonPolchinski,DouglasMoore}. 

Orbifold theories that correspond to projections of the
$\C{N}=4$ $\su{N}$ theory by a discrete subgroup $\Gamma$
of the $\su{4}$ R-symmetry group have been considered in  
Refs.~\cite{KachruSilver}--\cite{Gukov}. 
The resulting conformal field theories have either 
$\C{N}=0$, 1, or 2 supersymmetries, and the gauge group 
is generically of the form $\prod\limits_i\u{N_i}$. 
They are known to be dual to IIB string theory on
$AdS_5 \times S^5/\Gamma$.

The near-horizon description of conformal field theories
arising from D3-branes in orientifold backgrounds has been considered in 
Refs.~\cite{Kakus}--\cite{GremmKapustin}. 
The resulting field theories are dual to string theory 
on $AdS_5 \times S^5/G_{\rm orient}$, 
where the geometric part of the orientifold group $G_{\rm orient}$ 
is again a discrete subgroup of $\su{4}_R$.

In this paper,
we analyze the large $N$ supergravity descriptions 
of the type IIB models T-dual 
to a class of elliptic type IIA brane configurations.
This class of models contains two orientifold six-planes,
$N$ D4-branes, (possibly) D6-branes, and at most two NS 5-branes.
The T-dual IIB configurations contain
$N$ D3-branes in the background of an orientifold 7-plane 
and (possibly) a $\IZ_2$ orbifold and/or D7-branes.
The four-dimensional theories arising from the worldvolume dynamics 
of these configurations 
are $\C{N}=2$ (or $\C{N}=4$) superconformal gauge theories 
with simple gauge groups, or product groups with at most two factors.
(The supergravity descriptions of the models 
in Sect.~3.1, 3.2, and 3.4 of this paper
have been previously treated in 
Refs.~\cite{WittenBaryon}-\cite{GukovKapustin},
and are included here so that comparisons can be made
among this entire class of elliptic models.)

In Sect.~2, we review the details of the IIA and IIB configurations
for this class of theories.
The large $N$ supergravity description of these models
is the subject of Sect.~3, 
in which we identify the spectrum of relevant and marginal
chiral primary states belonging to each of these theories.
The AdS/CFT correspondence is used to match the chiral 
primary states of the supergravity description to
gauge invariant operators of the corresponding superconformal theories.

Seiberg-Witten curves \cite{SeibergWitten}
for the Coulomb branch of the elliptic models considered
in this paper have previously been obtained using M-theory methods 
\cite{WittenMtheory}--\cite{Elliptic}.
These constructions involve deformations of the $\C{N}=4$
superconformal field theories, in which the hypermultiplets are massive.
A direct connection between the Seiberg-Witten and AdS/CFT
descriptions of elliptic models is not yet evident,
although the AdS/CFT correspondence has recently been generalized 
to non-conformal theories arising from relevant deformations
\cite{PolchinskiStrassler,AharonyRajaraman}.

\newpage

\section{Description of the models}
\renewcommand{\theequation}{\thesection.\arabic{equation}}
\setcounter{equation}{0}
\subsection{Type IIA configurations}

We begin by reviewing the configurations in Type IIA string theory
that lead to orientifold elliptic models
\cite{WittenMtheory}--\cite{Elliptic}.
These configurations consist of 
NS 5-branes along 012345,
D4-branes along 01236, 
D6-branes along 0123789, 
and orientifold six-planes (O6-planes) parallel to the D6-branes. 
The D4-branes have finite extent along $x^6$, 
which is taken to be compact. 
This brane configuration
generically leaves $1/4$ of the original supersymmetries unbroken,
so the effective four-dimensional theory living on the 0123 directions 
of the D4-branes has $\C{N}=2$ supersymmetry.

Since the $x^6$ direction is compact, there are two O6-planes. 
As O6-planes carry $\pm 4$ units of RR charge 
(the corresponding planes will be denoted O$6^{\pm}$),
there are {\it a priori} three possibilities to consider:
O$6^{+}$--\,O$6^{+}$, 
O$6^{-}$--\,O$6^{-}$, 
and O$6^{+}$--\,O$6^{-}$. 
The first possibility cannot
be realized without breaking supersymmetry, as cancellation of RR charge in
the compact direction would require the introduction of anti D6-branes; 
we will not consider it further. 
The second possibility can be realized without breaking supersymmetry, 
which requires the presence of 4 D6-branes (plus mirrors) 
to ensure the cancellation of the RR charge;
this condition is equivalent to the finiteness of the
low-energy field theory on the D4-branes. 
Finally, in the third possibility,
the cancellation of RR charge 
(or, equivalently, finiteness of the field theory) 
requires the absence of D6-branes. 
In this paper, 
we will be studying two different classes of models, 
corresponding to the configurations 
O$6^{-}$--\,O$6^{-}$ (plus D6-branes) and
O$6^{+}$--\,O$6^{-}$. 

\vspace{1.0cm}
\noindent{\bf Spectra}

\noindent The spectra of the effective field theories realized 
on the above configurations
depends critically on the number and positions of the NS 5-branes. 
In this work we restrict ourselves to models with at most two NS 5-branes,
corresponding to elliptic models with simple gauge groups or product
gauge groups with no more than two factors.

\begin{figure}[hbtp]
\begin{center}
\mbox{\psfig{file=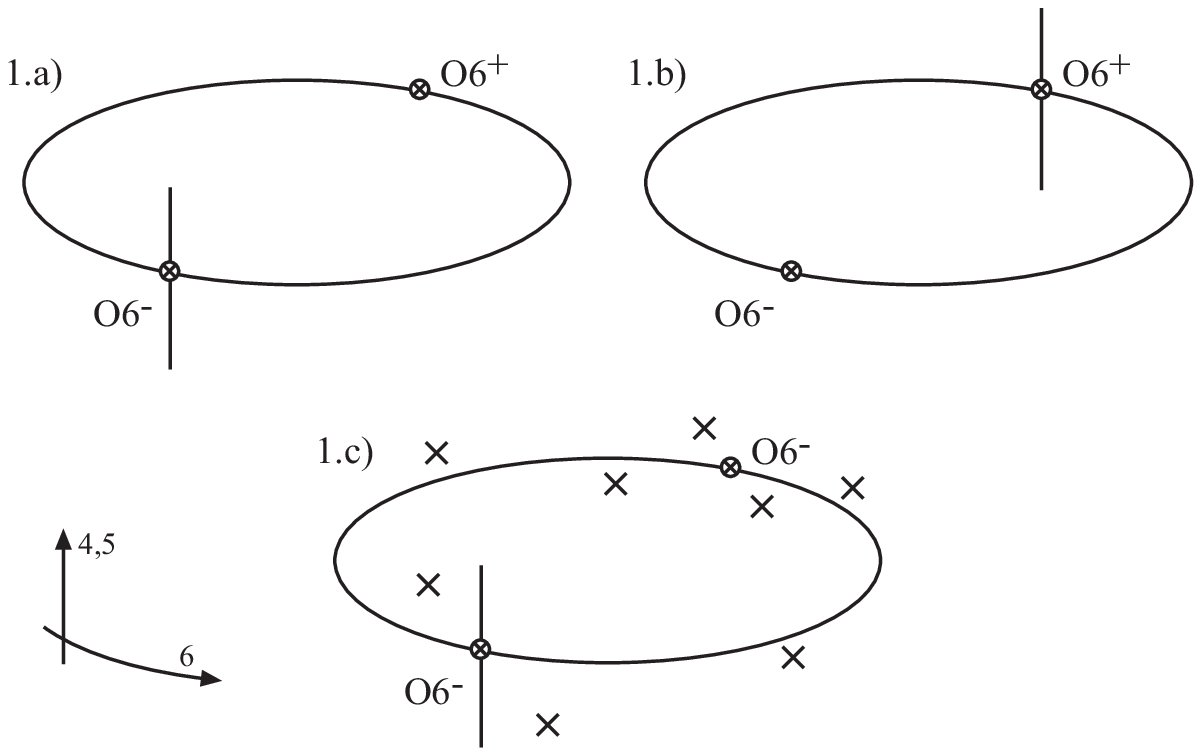,height=5.6cm}}
\end{center}
\end{figure}
{\vbox{\baselineskip12pt{\noindent\small{\bf Figure 1:} 
The three models with one NS 5-brane. 
The short vertical lines $|$ represent the NS 5-branes, 
the crossed circles $\otimes$ are the orientifold planes, 
while the crosses $\times$ denote the D6-branes, when present. 
D4-branes (not shown) extend along $x^6$.
The brane configurations must be invariant under the orientifold $\IZ_2$
symmetry inverting 456.\\ {\bf 1.a):}
$\so{N}$ + adjoint.
\\ {\bf 1.b):}
$\sp{2N}$ + adjoint.
\\ {\bf 1.c):} $\sp{2N}$ + $\Yasymm$ + 4$\fund$.
}}}

Figure 1 depicts the three possibilities for models with a single NS 5-brane.
The O$6^{+}$--\,O$6^{-}$ configurations in Figs.  1.a  and 1.b 
give rise to an $\C{N}=4$ gauge theory with gauge groups
$\so{N}$ and $\sp{2N}$  respectively,\footnote{Our convention is that
$\sp{2N}$ has rank $N$, so $\sp{2}\equiv \su{2}$.}
or equivalently, an $\C{N}=2$ vector multiplet coupled to a
massless hypermultiplet in the adjoint representation of the gauge group. 
The O$6^{-}$--\,O$6^{-}$ configuration in Fig. 1.c gives rise to an
$\C{N}=2$ gauge theory with gauge group $\sp{2N}$ 
coupled to a hypermultiplet in the antisymmetric representation
$\Yasymm$ and four hypermultiplets in the fundamental representation
$\fund$ of the gauge group.

\begin{figure}[hbtp]
\begin{center}
\mbox{\psfig{file=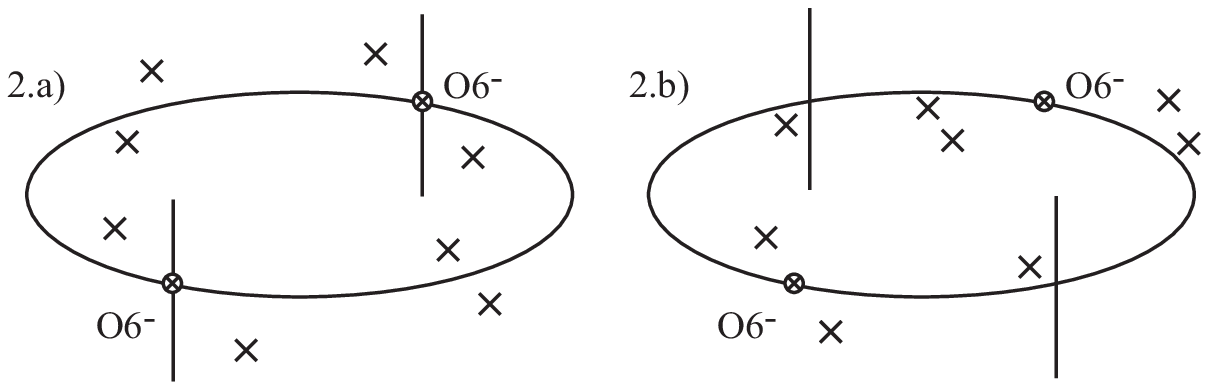,height=3cm}}
\end{center}
\end{figure}

{\vbox{\baselineskip12pt{\noindent\small{\bf Figure 2:} The
two models with two NS 5-branes in an O$6^{-}$--\,O$6^{-}$ background. \\
{\bf 2.a):} $\su{N}$ + 2$\Yasymm$ + $4\,\fund$. \\ 
{\bf 2.b):} $\sp{2N}_1\times\sp{2N}_2$ + 
$(\fund_1\,,\fund_2) + 2\,\fund_1 + 2\,\fund_2$. }}} 

Figure 2 depicts models with 2 NS 5-branes in an
O$6^{-}$--\,O$6^{-}$ background.
There are two possibilities, 
depending on whether the NS 5-branes intersect the O6-planes or not. 
If the NS 5-branes intersect the O6-planes, we have the model in
Fig. 2.a, which gives rise to an $\C{N}=2$ theory with gauge group
$\su{N}$, two hypermultiplets in the antisymmetric $\Yasymm$ representation
and four hypermultiplets in the fundamental $\fund$ representation. On the
other hand, if the NS 5-branes do not intersect the O6-planes (Fig. 2.b), we
obtain an
$\C{N}=2$ theory with gauge group $\sp{2N}\times\sp{2N}$, a hypermultiplet
in the bifundamental $(\fund\,,\fund)$ representation, and two
hypermultiplets in the fundamental representation $\fund$ of each of the two
factors. In this case the NS 5-branes are free to move along $x^6$ (while
respecting the symmetry $x^6\to -x^6$). This motion corresponds to a
marginal deformation changing the relative value of coupling constants of
the $\sp{2N}$  factors in the field theory on the D4-branes. 
Similarly, the  motion of the NS 5-brane in the (hidden)
eleventh direction $x^{10}$ changes the relative value of the theta angles
of the $\sp{2N}$  factors. (The average value of  the coupling constants and
theta angles is encoded in the M-theory description in the complex
structure of the torus defined by the compact directions $x^6$ and
$x^{10}$. Changing this complex structure -- that is, the relative
values of the radii $R_6$ and $R_{10}$ -- corresponds to a second pair of
marginal deformations in the field theory.)

\begin{figure}[hbtp]
\begin{center}
\mbox{\psfig{file=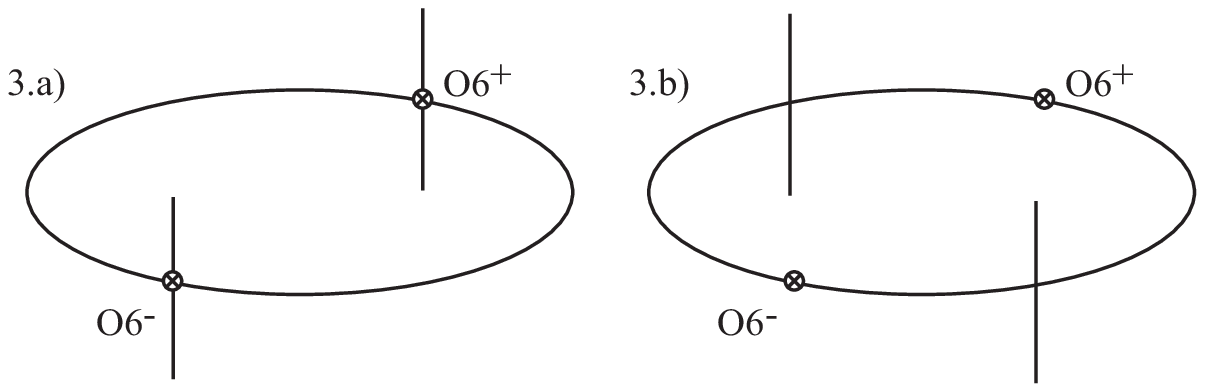,height=3cm}}
\end{center}
\end{figure}
{\vbox{\baselineskip12pt{\noindent\small{\bf Figure 3:} The
two models with two NS 5-branes in an O$6^{+}$--\,O$6^{-}$ background. \\
{\bf 3.a):} $\su{N}$ + $\Yasymm$ + $\Ysymm$. \\ 
{\bf 3.b):} $\sp{2N}\times\so{2N+2}$ + $(\fund\,,\fund)$.  }}}

Figure 3 depicts models with 2 NS branes in an
O$6^{+}$--\,O$6^{-}$ background.
As in the previous case, 
there are two possibilities to consider. The first one (Fig. 3.a)
gives rise to an $\C{N}=2$ theory with gauge group $\su{N}$, a
hypermultiplet in the symmetric representation $\Ysymm$, and a second 
hypermultiplet in the antisymmetric representation $\Yasymm$. 
In the second
possibility (Fig. 3.b) we obtain an $\C{N}=2$ theory with gauge group 
$\sp{2N}\times\so{2N+2}$ and a hypermultiplet
in the bifundamental $(\fund\,,\fund)$ representation,
where the ranks of the factor groups are fixed 
by the finiteness of the theory.
As in the
$\sp{2N}\times\sp{2N}$  case,  the NS 5-branes are free to move along the $x^6$ and
$x^{10}$ directions, the motions corresponding in the field theory to 
marginal deformations changing the relative couplings and theta angles of
the $\sp{2N}$  and $\so{2N+2}$ factors.

In all the cases discussed above, 
the beta functions for all the gauge groups vanish exactly.

\subsection{Type IIB configurations}
\la{2B}

The T-dual type IIB description of the previous models has been
systematically discussed by Park and Uranga \cite{ParkUranga} 
(see also Ref.~\cite{ErlichHanany}), 
so we will limit ourselves to a brief summary of
results, and refer the reader to the above papers for further details and
references.  
In the T-dual description, all the above models are realized as the
low-energy effective theory on a stack of  type IIB D3-branes probing a
certain mixed orbifold-orientifold background whose details depend on the
original type IIA model. As a general rule, T-duality along $x^6$ maps  NS
5-branes stretched along 012345 to a
$\IZ_k^{\rm orb}$ orbifold acting on 6789, where $k$ is the total number of
NS 5-branes (counting mirrors) present in the type IIA configuration. The
O6-planes become a single O7-plane, and the D6-branes, when present,
become IIB D7-branes. 

Let us now consider in turn the T-dual descriptions of the 
type IIA configurations shown in each of the figures above.

{\bf 1.a-b)} 
For the general case of $k$ NS 5-branes 
in an O$6^{+}$--\,O$6^{-}$ background, Park and Uranga 
\cite{ParkUranga}   (see also Ref.~\cite{DabolPark2}) 
have given the T-dual description as 
a stack of D3 branes probing 
an O7 background with the orientifold group
\be
G_{\rm orient} = \IZ_{k}^{\rm orb} + \IZ_{k}^{\rm orb}\alpha\Omega',
\la{orientigrup}
\ee
where the $\IZ_k^{\rm orb}$ generator $\theta$ acts on $z_1 = x^6 +i x^7$
and 
$z_2 = x^8+ix^9$ as
\be
z_1\buildrel \theta\over\to  {\rm e}^{2\pi i/k} z_1, \qquad
z_2\buildrel \theta\over\to {\rm e}^{-2\pi i/k} z_2,
\la{orbifold}
\ee
and $\alpha^2=\theta$. Also, $\Omega'=\Omega R_{45}(-1)^{F_L}$, where
$\Omega$ is the world-sheet parity operation, 
$R_{45}$ is a reflection in 45, and $(-1)^{F_L}$ changes the sign of the
Ramond sector of left-movers. 
Orientifold projections of this kind have
been studied in Refs.~\cite{DabolPark2}-\cite{WittenToroidal}. 

In figures 1.a and 1.b, there is only a single NS 5-brane ($k=1$),
so that the $\IZ_k^{\rm orb}$ orbifold is trivial ($\theta =1$).
However, $\alpha = \theta^{1/2} \equiv R_{6789}$, 
so eq.~\eqs{orientigrup} becomes
\be
G_{\rm orient} = \{\identity , R_{6789}\Omega'\}=\{\identity,\Omega
R_{456789}(-1)^{F_L}\}.
\la{orientin4}
\ee
This is actually the orientifold group of an O3-plane extending along 0123
(parallel to the probe D3-branes),  
precisely the setup considered in Refs.~\cite{Kakus,WittenBaryon}. 
Although the two
different IIA configurations in figs 1.a and 1.b
appear to give rise to the same IIB configuration, 
these configurations are
actually distinguished by the background values of the 
two-form fields \cite{WittenBaryon}.   

{\bf 1.c)} 
The T-dual of the configuration with $k$ NS 5-branes in an
O$6^{-}$--\,O$6^{-}$ background contains D3-branes probing an O7-plane
background with the orientifold group
\be
G_{\rm orient} = \IZ_{k}^{\rm orb} + \IZ_{k}^{\rm orb}\Omega',
\la{orientigrupii}
\ee
with $\IZ_k^{\rm orb}$ and $\Omega'$  as defined above \cite{ParkUranga}.
Orientifold projections of this kind have been considered in
\cite{GimonPolchinski,DabolPark2,GimonJohnson,BlumIntri1} for the
realization of six-dimensional theories.

In figure 1.c, $k=1$, so the $\IZ_k^{\rm orb}$ orbifold is again trivial,
yielding the orientifold group
\be
G_{\rm orient} = \{\identity , \Omega R_{45}(-1)^{F_L}\},
\la{orientispanti}
\ee
which corresponds to an O7-plane extending along 01236789. The background
also contains 4 D7-branes (plus mirrors) parallel to the O7-plane, which
give rise to the fundamental hypermultiplets of the field theory on the
probe D3-branes. 

{\bf 2.a-b)} 
Figure 2 contains two NS 5-branes, 
so that the T-dual configuration consists of a stack of D3-branes in the
background of the orientifold \eqs{orientigrupii} with $k=2$:
\be
G_{\rm orient} = \IZ_{2}^{\rm orb} + \IZ_{2}^{\rm orb}\Omega',
\la{orientiminusminus}
\ee
where $\IZ_{2}^{\rm orb}$ acts as $(z_1,z_2)\to (-z_1,-z_2)$, with $z_1$,
$z_2$ as above. The corresponding O7-plane extends along 01236789, and the
background also includes 4 D7-branes (plus mirrors). Again we find that two
different IIA configurations lead  apparently to the same IIB configuration.
In fact, the IIB descriptions of models 2.a and 2.b are distinguished 
by the action of the orientifold projection $\Omega'$ on the
twisted sectors of the $\IZ_2^{\rm orb}$ orbifold,
and by the action of the $\IZ_2^{\rm orb}$ orbifold projection 
on the D7 brane states. 
As pointed out in Ref.~\cite{ParkUranga}, 
and previously discussed in connection with Type I and
six-dimensional IIB models T-dual to those considered here
\cite{BlumIntri1,BrunnerKarch2,HananyZaffa},
$\Omega'$ acts with an additional minus sign on the $\IZ_2^{\rm orb}$
twisted sector of model 2.b (corresponding, in the Type I context, to
models {\it with vector structure}; the other possibility, leading to model
2.a, being related in the Type I context to models {\it without vector
structure} \cite{WittenToroidal}.)   
Equivalently, there are two alternative
(and inequivalent) choices for the Chan-Paton factors, 
each of which corresponds, in the type IIB context, 
to each of the Type IIA configurations \cite{ParkUranga}. 

{\bf 3.a-b)} 
In figure 3, there are two NS branes ($k=2$), 
so the orientifold group \eqs{orientigrup} becomes 
\be
G_{\rm orient} = \IZ_{2}^{\rm orb} + \IZ_{2}^{\rm orb}\alpha\Omega',
\la{orientiplusplus}
\ee
where $\IZ_2^{\rm orb}=\{\identity, \alpha^2\}$ with $\alpha^2 = R_{6789}$. 
As in the previous case, both IIA configurations lead to the same orientifold
projection in the Type IIB description, but exactly as above, $\Omega'$
acts with an additional minus sign on the $\IZ_2^{\rm orb}$ twisted sector
of model 3.b.  

\vbox{
\begin{center}
\begin{tabular}{||c|c|c||}
\hline\hline
 type IIA & type IIB  & Gauge theory  
\\[-.1in]
{} & Orientifold Group & {}
\\
\hline\hline
O$6^+$--\,O$6^-$ with 1 NS5 & {} & $\sp{2N}\,+\,$ adjoint
\\[-.15in]
{} & $\{\identity , R_{6789}\Omega'\}$
& or
\\[-.15in] 
{\footnotesize\bf (1.a-b)} & 
$=\{\identity,\Omega R_{456789}(-1)^{F_L}\}$ 
&  $\so{N} \,+\,$ adjoint
\\
\hline
O$6^-$--\,O$6^-$ with 1 NS5 & {} & {}
\\[-.15in]
{} & $\{\identity , \Omega R_{45}(-1)^{F_L}\}$ & $\sp{2N}\,+$ 
$\Yasymm$\,\,+ $4\,\fund$
\\[-.15in]
{\footnotesize\bf (1.c)} & {} & {}
\\
\hline
O$6^-$--\,O$6^-$ with 2 NS5 & {} &
$\su{N}\,+\,$ 2 $\Yasymm$ $+$ $4\,\fund$
\\[-.15in]
{} & $\IZ_{2}^{\rm orb} + \IZ_{2}^{\rm orb}\Omega'$ & or 
\\[-.15in]
{\footnotesize\bf (2.a-b)} & {} & 
$\sp{2N}_1\times\sp{2N}_2+ 
(\fund{}_{1},\fund{}_{2})+2\, \fund\,{}_{1}+ 2\, \fund\,{}_{2}$
\\
\hline
O$6^+$--\,O$6^-$ with 2 NS5 &  & $\su{N}$ + $\Yasymm$ + $\Ysymm$
\\[-.15in]
{} & $\IZ_{2}^{\rm orb} + \IZ_{2}^{\rm orb}\alpha\Omega'$ & or 
\\[-.15in] 
{\footnotesize\bf (3.a-b)} & {} & 
$\sp{2N} \times \so{2N+2}+ (\,\fund\,,\fund\,)$ 
\\ 
\hline\hline
\end{tabular}
\end{center}
\centerline{{\footnotesize{\bf Table {\tablecomp}}: 
Comparative geometries. }}} 

\newpage
\section{Supergravity descriptions in the large $N$ limit}
\setcounter{equation}{0}

When the number of probe D3-branes is large, 
the superconformal theories propagating in the world-volume 
of the brane configurations 
are conjectured to be dual to type IIB supergravity 
(or string theory)
on a space containing an $AdS_5$ factor.
The original formulation of the conjecture
\cite{MaldacenaAds}--\cite{AdsReview}
related the
$\C{N}=4$ $\su{N}$ gauge theory to type IIB supergravity on $AdS_5\times S^5$. 
The spectrum of type IIB supergravity on $AdS_5\times S^5$ 
falls into representations of the $\so{6}$ isometry group
of $S^5$ \cite{Sugra}, 
which corresponds to the $\su{4}_R$ $R$-symmetry of the $\C{N}=4$ theory. 
There are three scalar families in the supergravity spectrum 
containing states with zero or negative mass, summarized in 
Table  {\tablescalars}
(adapted from Ref.~\cite{Ansar1}):
\vspace{.1in}

\vbox{
\begin{center}
\begin{tabular}{||l|c|c|c|c|l||}
\hline\hline
$\Delta$ & $m^2$ & Range & Dynkin Label & 
$\su{4}_R$ & Type IIB origin
\\
\hline\hline
$k$ & $k(k-4)$ & $k\geq 2$ & $(0,k,0)$ & ${\bf 20'}$, ${\bf 50}$, ${\bf
105}$, \ldots & graviton + 4-form
\\  
$k+3$ & $(k-1)(k+3)$ & $k\geq 0$ & $(0,k,2)$ & ${\bf 10}$, ${\bf 45}$, 
${\bf 126}$, \ldots & 2-forms
\\ 
$k+4$ & $k(k+4)$ & $k\geq 0$ & $(0,k,0)$ & ${\bf 1}$, ${\bf 6}$, 
${\bf 20'}$, \ldots & dilaton/axion 
\\ 
\hline\hline
\end{tabular}
\end{center}
\begin{center}
{\footnotesize{\bf Table \tablescalars}: 
Scalar families of IIB supergravity on $AdS_5\times S^5$ 
containing $m^2\leq 0$ states \cite{Sugra}.} 
\end{center}
}
\noindent 
Here $m^2 = (\Delta - p)(\Delta +p -d)$ for a $p$-form state on $AdS_{d+1}$
with $AdS$ mass $m$ coupling (in the sense of Witten \cite{WittenAdS}) to an
operator of dimension
$\Delta$ of the boundary $\C{N}=4$ theory. 
There is also one family of vector fields 
that contains massless states 
(at level $k=1$ in Table \tablevectors). 

\vbox{
\begin{center}
\begin{tabular}{||c|c|c|c|c|c||}
\hline\hline
$\Delta$ & $m^2$ & Range & Dynkin Label & 
$\su{4}_R$ & Type IIB origin
\\
\hline\hline
$k+2$ & $(k-1)(k+1)$ & $k\geq 1$ & $(1,k-1,1)$ & ${\bf 15}$, ${\bf 64}$, 
${\bf 175}$, \ldots & graviton + 4-form
\\  
\hline\hline
\end{tabular}
\end{center}
\begin{center}
{\footnotesize{\bf Table \tablevectors}: 
Vector families of IIB supergravity on
$AdS_5\times S^5$ containing $m^2=0$ states\cite{Sugra}.} 
\end{center}
}

\noindent These $k=1$
massless vector states transform in the adjoint of $\su{4}_R$ and couple to
the $\su{4}_R$ $R$-symmetry currents (with $\Delta=3$) of the $\C{N}=4$ theory.

In section 2, we described a number of $\C{N}=4$ or $\C{N}=2$ 
superconformal theories propagating in the world-volume 
of brane configurations involving orientifold planes.
These theories are conjectured to be dual to type IIB supergravity 
on $AdS_5\times S^5/G_{\rm orient}$, 
where the orientifold group $G_{\rm orient}$ depends on the specific model
\cite{Kakus}--\cite{GremmKapustin} \cite{AdsReview}. 
The spectrum of this supergravity background will include
(i) the KK reduction of ten-dimensional type IIB theory 
on $AdS_5\times S^5/G_{\rm orient}$
(obtained by a $G_{\rm orient}$ projection of the 
$AdS_5\times S^5$ states), 
and 
(ii) states (invariant under $G_{\rm orient}$)
in sectors supported on $AdS_5\times Y$,
where $Y\subset S^5/G_{\rm orient}$
is the fixed-point set of $G_{\rm orient}$.
If the fixed-point set is empty 
(as, for example, in the $\C{N}=4$ $\so{N}$ and $\sp{2N}$  theories), 
the spectrum of the theory only contains states in (i),
and is a subset of the $AdS_5\times S^5$ spectrum.

An alternative approach \cite{Ansar2},
which we will not follow in this paper,
is to derive the part of the spectrum coming from the
$AdS_5$ supergravity modes by diagonalizing 
the linearized equations of motion of the supergravity fields 
on the compact space $S^5/G_{\rm orient}$. 
The equations for the scalar modes turn out to be 
Laplace equations on $S^5/G_{\rm orient}$, 
easily written down once the corresponding metric is known. 
In all the cases considered in this paper, 
the near-horizon limit of the D3-brane metric is similar to the
$AdS_5\times S^5$ solution 
(see for example \cite{Ansar1}, Eq. (13)), 
but with the metric 
$d\widetilde{\Omega}_5{}^2$ of the compact space $S^5/G_{\rm orient}$ 
given by the angular part of 
\be
ds^2_{S^5/G_{\rm orient}} = |dz_1|^2 + |dz_2|^2 + |dw|^2=dr^2 + r^2d\widetilde{
\Omega}_5{}^2,
\la{metric}
\ee
where the coordinates $z_1 = x^6+ix^7$, $z_2 =x^8+ix^9$, $w = x^4+ix^5$ are
subject to various identifications which depend on the form 
of $G_{\rm orient}$.

\subsection{$\sp{2N}$ + adjoint, and $\so{N}$ + adjoint}

Let us start by reviewing the analysis of the  $\C{N}=4$ $\so{N}$ and
$\sp{2N}$ theories, or equivalently, 
$\C{N}=2$ $\so{N}$ and
$\sp{2N}$ theories with massless adjoint hypermultiplets.
As we saw in Sect.~2.2, 
the orientifold group $G_{\rm orient}$ is that of an O3-plane 
acting on the $\IR^6$ transverse to the D3-branes as 
\be x^{4,5,6,7,8,9}\to -x^{4,5,6,7,8,9}.
\la{spso}
\ee 
In the near horizon limit the geometry becomes 
$AdS_5 \times S^5/\IZ_2$, where the
geometric part of the $\IZ_2$ action on $S^5$ is the same as the action of
\eqs{spso} on the angular part of the transverse $\IR^6$. 
Since $S^5$ with such a
$\IZ_2$ action is actually $\IR{\rm P}^5$, 
the $\sp{2N}$  and $\so{N}$ theories are dual to string
theory on $AdS_5 \times \IR {\rm P}^5$ \cite{WittenBaryon}. Further, the
$\IZ_2$ action has no fixed points on $S^5$, so the supergravity spectrum
is simply a truncation of that for the $\C{N}=4$ $\su{N}$ theory,
which can be easily determined as follows.  
The isometry group of $S^5$ is $\so{6}\simeq\su{4}_R$, which
is also the
$R$-symmetry group of the $\C{N}=4$ $\su{N}$ theory. The coordinates
$x^{4,5,6,7,8,9}$ transform in the $\rep6$ of $\su{4}_R$, so the $\IZ_2$
action
\eqs{spso} is actually a $\IZ_4$ in the center of $\su{4}_R$ 
(it acts as $i\identity$ on the $\rep4$). 
In addition, the
$\Omega (-1)^{F_L}$ piece of the orientifold group acts with an additional
minus sign on the states coming from the ten-dimensional second-rank
antisymmetric tensors, leaving all other states invariant. 
Therefore, the spectrum consists
of those states in Table {\tablescalars}
invariant under the combination of
$\IZ_4$ and $\Omega (-1)^{F_L}$,
which are listed in Table {\tblspso} .
\vbox{
\begin{center}
\begin{tabular}{||l|c|c|c|l||}
\hline\hline
$\Delta$ & $m^2$ & Range & $\su{4}_R$ & CFT Operator
\\
\hline\hline
$2k$ & $4k(k-2)$ & $k\geq 1$   & ${\bf 20'}$, ${\bf 105}$, \ldots 
& tr\,$(\phi^{\{I_1}\cdots\phi^{I_{2k}\}})
$
\\  
$2k+3$ & $(2k-1)(2k+3)$ & $k\geq 0$   & ${\bf 10}$, ${\bf 126}$, \ldots 
& $\epsilon^{\alpha\beta}$\,tr\,$(\lambda_{\alpha
A}\lambda_{\beta B}\,\phi^{\{I_1}\cdots \phi^{I_{2k}\}})+\cdots$
\\ 
$2k+4$ & $4k(k+2)$ & $k\geq 0$ &   ${\bf 1}$,  ${\bf 20'}$, \ldots 
& tr\,$(F^2_{\mu\nu}\phi^{\{I_1}\cdots\phi^{I_{2k}\}})+\cdots$
\\ 
\hline\hline
\end{tabular}
\end{center}
\begin{center}
\begin{minipage}{5.9in}
\baselineskip12pt
{\noindent\small{\bf Table {\tblspso}}: 
Scalars in chiral primary representations of $\C{N}=4$\,
$\sp{2N}$ and $\so{N}$ \cite{WittenBaryon,AdsReview}.  \\
$I_j$ is an index in the ${\bf 6}$ of $\su{4}_R$,
and tr\,$(\phi^{\{I_1}\cdots\phi^{I_{2k}\}})$ denotes the
symmetric traceless \\
product of ${\bf 6}$'s; {\it e.g.}, 
tr\,$(\phi^{\{I}\phi^{J\}})
={\rm tr}(\phi^I\phi^J)-\frac{1}{6}\delta^{IJ}{\rm tr}(\phi^K\phi^K).$
}
\end{minipage}
\end{center}
}

\subsection{$\sp{2N}$ + 1 antisymmetric + 4 fundamental hypermultiplets}
\la{sectspanti}

The $\C{N}=2$ $\sp{2N}$ theory
with 1 antisymmetric and 4 fundamental hypermultiplets
(all massless)
arises as the effective theory on $N$ D3-branes 
probing the F-theory background first discussed by Sen \cite{Sen1}. 
This model has been much studied 
both from the field theory 
\cite{BanksDouglasSeiberg}--\cite{DouglasLoweSchwarz} 
and dual supergravity \cite{Ansar1,Ansar2} viewpoints. 
The analysis of the AdS/CFT correspondence 
in Refs.~\cite{Ansar1,Ansar2,ParkUranga} 
is already very complete, 
so we limit ourselves to a brief review of the results.

In Sect.~\ref{2B} we found the orientifold group to be 
\be
G_{\rm orient} = \{\identity , \Omega R_{45}(-1)^{F_L}\}
\simeq \IZ_2^{\rm orient},
\la{orientispantitwo}
\ee
where the geometric part of $\IZ_2^{\rm orient}$ 
acts on $w = x^4+ix^5$ as $w\to-w$. 
The $\IZ_2^{\rm orient}$ action fixes the hyperplane $w=0$ 
(which corresponds to the classical positions of the O7-plane 
and the D7-branes for the theory with massless fundamentals) 
in the six-dimensional  space transverse to the D3-branes,
which becomes $S^3\subset S^5$ in the near-horizon geometry.

In the near-horizon limit the metric for the probe D3-branes
looks like $AdS_5\times S^5/\IZ_2^{\rm orient}$, with the metric of the
compact space given by 
\be
&&d
\widetilde{
\Omega}_5{}^2 = d\theta_1{}^2 + 
\cos^2\theta_1 \,\left(d\theta_2{}^2 +
\cos^2\theta_2\,d\phi_1{}^2+\sin^2\theta_2\,d\phi_2{}^2\,\right)+
\sin^2\theta_1\,
d\phi_3{}^2,\ret &&\qquad\theta_{1,2}\in [0,\pi/2],\qquad\qquad
\phi_{1,2}\in [0,2\pi],\qquad\qquad \phi_{3}\sim \phi_{3}+\pi, 
\la{metricspanti}
\ee
where we have used the following parametrization:
\be
z_1=r\cos{\theta_1}\cos{\theta_2}\,e^{i\phi_1}, \qquad 
z_2=r\cos{\theta_1}\sin{\theta_2}\,e^{i\phi_2}, \qquad 
w=r\sin{\theta_1}\,e^{i\phi_3}.
\la{param}
\ee

Notice that the $\IZ_2^{\rm orient}$
orientifold action \eqs{orientispantitwo}
differs from that in the $\C{N}=4$
$\sp{2N}$ /$\so{N}$ case discussed in the previous subsection 
(so in this case $S^5/\IZ_2^{\rm orient} \not \simeq \IR {\rm P}^5$). 
More precisely, 
the isometry group of $S^5$ is $\so{6} \supset  \so{4} \times \so{2}$,
where 
$\so{4} \simeq \su{2}_L\times\su{2}_R$ 
is the rotation group on $\IC^2\sim \{z_1, z_2\}$, 
and 
$\so{2} \simeq \u{1}_R$ 
is the rotation group on $\IC\sim \{w\}$.  
The geometric part of the orientifold group \eqs{orientispantitwo}
acts only on $\u{1}_R$,
identifying $e^{i\phi_3}$ with $e^{i(\phi_3+\pi)}$. 
On the field theory side, 
$\su{2}_R\times\u{1}_R/\IZ_2^{\rm orient}$ is the 
$\C{N}=2$ $R$-symmetry group, 
while $\su{2}_L$ is the flavor group of the
hypermultiplet in the antisymmetric representation,
as we will explain further below.

The supergravity spectrum consists of 
(a) the $\IZ_2^{\rm orient}$ projection of the $AdS_5\times S^5$ states,
plus (b) the contribution from the fixed-point set $S^3 \subset S^5$ of
$\IZ_2^{\rm orient}$, 
which is actually the world-volume theory of 8 D7-branes wrapping
$S^3$ and filling the $AdS_5$ factor. 

\vs{.1in}
\noindent{\large\bf (a) Bulk sector:}
\vs{.1in}

To find the $\IZ_2^{\rm orient}$--invariant states 
of the $AdS_5\times S^5$ spectrum,
one decomposes the $\su{4}_R$ representations 
in Table {\tablescalars} into representations 
$(d_L, d_R)_{q_R}$ of $\su{2}_L\times\su{2}_R\times\u{1}_R$,
where $d_L$ and $d_R$ are the dimensions of the $\su{2}_L$ and $\su{2}_R$
representations, 
and $q_R$ is the $\u{1}_R$ charge of the state,
normalized such that $w$ has $q_R=2$.
(Table {\tblbranching} lists the relevant branching rules.)
Each state gets a phase $e^{i\pi q_R/2}$ under $G_{\rm orient}$, 
so the invariant states for representations in the
first and third lines of Table {\tablescalars}
are those with $q_R = 0 \mod4$.
The orientifold acts with an additional minus sign 
on states in the second line of Table {\tablescalars}
because of the action of $\Omega (-1)^{F_L}$ on the
ten-dimensional two-form fields. 
Thus, for these states,  the selection rule is $q_R = 2 \mod4$. 
Of the surviving states, 
only those satisfying the relation \cite{Minwalla}
\be
\Delta = \half \Big(\,\vert q_R\vert + 2(d_R -1)\,\Big),  
\la{chiralprimaries}
\ee
where $\Delta$ is the dimension of the corresponding operator in the CFT, 
are (dual to) chiral primaries of the conformal field theory. 
The results are summarized in Table {\tblspanti}.

\vbox{
\begin{center}
\begin{tabular}{||l|l||}
\hline\hline
 $\su{4}_R$ & $\su{2}_R\times\su{2}_L\times\u{1}_R$
\\
\hline\hline
 $\fund$ (\rep{4}) & $(2,1)_{\pm 1}$ 
\\
$\Yasymm$ (\rep{6}) & $(1,1)_{\pm2}  + (2,2)_0$ 
\\
$\Ysymm$ (\rep{10}) & $(1,3)_2 +(3,1)_{-2} +(2,2)_0$
\\
$\tableau{2 1 1}$ (\rep{15}) & $(1,1)_0 +(3,1)_{0} +(1,3)_0 +(2,2)_{\pm 2} $
\\
$\tableau{2 2}$ (\rep{20'}) & $(3,3)_0+(2,2)_{\pm 2}+(1,1)_{\pm4}+(1,1)_0$
\\
$\tableau{3 1}$ (\rep{45}) & $(4,2)_{-2}+(2,4)_2+(2,2)_{\pm2}+
(3,1)_{-4}+(1,3)_4+(3,3)_0+(3,1)_0+(1,3)_0$
\\
$\tableau{3 3}$ (\rep{50}) & $(4,4)_0+(3,3)_{\pm 2}+(2,2)_{\pm
4}+(2,2)_0+(1,1)_{\pm 6}+(1,1)_{\pm 2}$
\\
$\tableau{4 4}$ (\rep{105}) &
$(5,5)_0+(4,4)_{\pm2}+(3,3)_{\pm4}+(3,3)_0+(2,2)_{\pm6}+(2,2)_{\pm2}
+(1,1)_{\pm8}+(1,1)_{\pm4}+(1,1)_0$
\\ 
\hline\hline
\end{tabular}
\end{center}

\centerline
{\small{\bf Table {\tblbranching}}: Branching rules for
$\su{4}_R\to\su{2}_R\times\su{2}_L\times\u{1}_R$.
}
}
\vs{.1in}

\noindent {\large\bf (b) D7-brane sector:}
\vs{.1in}

\noindent The spectrum of chiral primaries coming from the D7-branes 
consists of the KK reduction of the eight-dimensional 
$\C{N}=1$ vector multiplet \cite{Ansar2}, 
subject to the constraint \eqs{chiralprimaries}. 
These states form representations of the gauge group on the D7-branes, 
which becomes the flavor symmetry group $\C{G}_7$ 
of the hypermultiplets in the
fundamental representation of the four-dimensional gauge group. 
In the present case $\C{G}_7 = \so{8}$, as appropriate for an $\sp{2N}$ theory
with four  hypermultiplets in the fundamental representation. 
We postpone the detailed analysis of the D7-brane sector until 
we study the $\su{N}+2\Yasymm+4\fund$ theory, 
and will just include the resulting chiral primary states
in Table {\tblspanti}.

\vbox{
\begin{center}
\begin{tabular}{||l|c|c|c|c|c|c||}
\hline\hline
$\Delta$ & Sector & $\su{4}_R$ & 
$(d_L, d_R)_{q_R}$ & 
$\so{8}_F$ & $m^2$ &  CFT operator  
\\
\hline\hline
2 & bulk & $\rep{20'}$ & $(1,1)_4$ &--& $-4$ & tr\,$\Phi^2$ 
\\  
2 & bulk & $\rep{20'}$ & $(3,3)_0$ &--& $-4$ & tr\,$[A_{(i}JA_{j)}J]$ 
\\ 
2 & D7 & & $(1,3)_0$ & 28& $-4$ & $Q^{[I}JQ^{K]}$
\\ 
3 & bulk & \rep{10} & $(1,3)_2$ &--& $-3$ & tr\,$\lambda^2$ 
\\ 
3 & bulk & \rep{50} & $(4,4)_0$ &--& $-3$ & tr\,$[A_{(i}JA_{j}JA_{k)}J]$ 
\\ 
3 & bulk & \rep{50} & $(2,2)_4$ &--& $-3$ & tr\,$[A_{i}J\Phi^2]$   
\\
3 & D7 &  & $(2,4)_0$ & $28$ & $-3$ & $Q^{[I}JA_{i}JQ^{K]}$ 
\\  
4 & bulk & \rep{105} & $(1,1)_8$ &--& 0 & tr\,$\Phi^4$ 
\\ 
4 & bulk & \rep{105} & $(3,3)_4$ &--& 0 & tr\,$[A_{(i}JA_{j)}J\Phi^2]$ 
\\ 
4 & bulk & \rep{105} & $(5,5)_0$ &--& 0 & tr\,$[A_{(i}JA_{j}JA_{k}JA_{l)}J]$  
\\ 
4 & bulk & \rep{45} & $(2,4)_2$ &--& 0 &
tr\,$[\lambda^2 A_iJ]$  
\\ 
4 & D7 & &  $(3,5)_0$ & 28 &0 & $Q^{[I}JA_{(i}JA_{j)}J Q^{K]}$ 
\\
\hline\hline
\end{tabular}
\end{center}
\begin{center}
\begin{minipage}{5.3in}
\baselineskip12pt
{\noindent\small{\bf Table {\tblspanti} }: 
Scalars with $\Delta \le 4$ in chiral primary representations of
$\sp{2N}$\,\,w/\,\,$\Yasymm\,\oplus\,4\,\fund$.
$d_R$, $d_L$ denote the dimensions of the
$\su{2}_R$, $\su{2}_L$ representations; 
$q_R$ is the $\u{1}_R$ charge;
$i,j$ are $\su{2}_L$ indices, and 
$I,J$ are $\so{8}_F$ indices.}
\end{minipage}
\end{center}
}

The last column of Table {\tblspanti} lists
the CFT operators that correspond to the 
supergravity states \cite{Ansar1, Ansar2}.
These operators are written as products of fields of the
perturbative $\sp{2N}+\Yasymm $ worldvolume theory,
which contains an $\C{N}=2$ vector multiplet in the adjoint representation,
and one antisymmetric and four fundamental $\C{N}=2$ hypermultiplets. 
Each of these $\C{N}=2$ multiplets consists of   
several $\su{2}_R$ representations, 
with definite scaling dimension $\Delta$.
For example, the vector multiplet
contains the scalar field $\Phi$, and the vector field $A_\mu$,
both in $\su{2}_R$ singlets,
and the Weyl fermions $(\lambda, \psi)$ in an $\su{2}_R$ doublet.
Those $\su{2}_R$ multiplets comprising the chiral primary operators 
in Table {\tblspanti} are listed in Table {\tblspantifield}.

\vbox{
\begin{center}
\begin{tabular}{||l|c|c|c|c|c||}
\hline\hline
$\Delta$ & $\sp{2N}$ & 
$\su{2}_L \times \su{2}_R \times \u{1}_R $ & 
$\so{8}_F$ & CFT field 
\\
\hline\hline
${3\over 2}$ &  adjoint & $(1,2)_1$ &  1 & $\lambda$
\\  
{1} & adjoint & $(1,1)_2$ & 1  &  $\Phi$
\\  
{1} & $\Yasymm$ & $(2,2)_0$ & 1 &  $A_i$
\\  
{1} & $\fund$  & $(1,2)_0$ & 8  &  $Q^I$
\\  
\hline\hline
\end{tabular}
\end{center}
\begin{center}
\baselineskip12pt
{\noindent\small{\bf Table {\tblspantifield} }: 
$\su{2}_R$ multiplets in perturbative
$\sp{2N}$\,\,w/\,\,$\Yasymm\,\oplus\,4\,\fund$}
\end{center}
}
\vspace{.5cm}

In all our tables, 
only the indices for the $\su{2}_L$ and $\so{8}_F$ symmetries
are shown explicitly;
the space-time,
gauge, 
and $\su{2}_R$ indices are implicit.
The chiral primary operators are always completely symmetric
in $\su{2}_R$ indices, 
and only the highest component of the $\su{2}_R$ multiplet
is exhibited (e.g., $\lambda$ for the doublet $(\lambda, \psi)$).

The flavor symmetry of the theory is $\su{2}_L \times \so{8}$,
where $\su{2}_L$ acts only on the antisymmetric hypermultiplet,
and $\so{8}$ only on the four fundamental hypermultiplets.
To understand this, recall that
$\sp{2N}$ has only real and pseudo-real representations, 
which are therefore self-conjugate. 
If $v_a$ transforms in the fundamental
representation of $\sp{2N}$, 
then $\bar v^a =J^{ab}v_b$ transforms in the 
conjugate (anti)fundamental representation,
where $a,b$ are $2N$-valued $\sp{2N}$  indices
and $J^{ab}$ is the invariant antisymmetric tensor of $\sp{2N}$.
$J^{ab}$ and its inverse, $J_{ab}$, are 
used to raise and lower $\sp{2N}$  indices.
An $\C{N}=2$  hypermultiplet in the antisymmetric
representation of $\sp{2N}$ consists of an $\C{N}=1$  chiral superfield
$A_{ab}=-A_{ba}$ in the antisymmetric representation, 
and a second chiral superfield
$\tilde A^{ab}=-\tilde A^{ba}$ in the conjugate antisymmetric
representation. 
The piece of the $\C{N}=2$ superpotential involving these
fields is 
\be
\tilde A^{ab}\Phi_b{}^c A_{ca},
\la{superman}
\ee
where $\Phi_b{}^c$ is a chiral superfield in the adjoint of $\sp{2N}$. 
This
superpotential has an obvious  $U(1)_L$ symmetry acting as 
$A\to e^{i\alpha}A$, $\tilde A\to e^{-i\alpha}\tilde A$, $\Phi\to\Phi$, 
which is
the expected flavor symmetry of a single hypermultiplet.  However, as the
antisymmetric representation of $\sp{2N}$  is self-conjugate, we can actually 
define two chiral superfields 
$A_{1 ab}\equiv A_{ab}$ and
$A_{2 ab}\equiv J_{ac}J_{bd} \tilde A^{cd}$, both transforming in the
antisymmetric representation. In terms of these new fields the
superpotential reads
\be
J^{ab}A_{2 bc}\widehat\Phi^{cd} A_{1 da},
\la{superman2}
\ee
where $\widehat\Phi^{cd}=J^{ac}\Phi_a{}^d$ is symmetric in $c,d$, 
as appropriate for the adjoint representation of $\sp{2N}$. 
Equation \eqs{superman2} is odd under exchange of $1\leftrightarrow 2$, 
so the superpotential actually possesses an $\sp{2}\simeq \su{2}$ symmetry
acting on the doublet $(A_1, A_2)$, which we can identify with  the
$SU(2)_L$ flavor symmetry that emerges from the geometry. 

A similar
argument holds for hypermultiplets in other representations, as follows:
given $N_k$ hypermultiplets transforming in a $k^{th}$ rank  
tensor representation of $\sp{2N}$, the flavor symmetry is enhanced to
$\sp{2N_k}$ if $k$ is even, and to $\so{2N_k}$ if $k$ is odd (in the latter
case, the superpotential turns out to be {\it symmetric} under exchange of
any pair of chiral superfields). For example, if we have 
$N_f$ hypermultiplets in the fundamental representation of $\sp{2N}$, the
flavor group is $\so{2N_f}$. This result 
is actually an extension to the higher-rank case of what happens in the
$\C{N}=2$ $\su{2}+ N_f\fund\simeq\sp{2}+ N_f\fund$ theory discussed by
Seiberg and Witten in the second paper of \cite{SeibergWitten}.

\subsection{$\su{N}$ + 2 antisymmetric + 4 fundamental hypermultiplets} 
\la{sect2anti}

The $\C{N}=2$ $\su{N}$ theory
with 2 antisymmetric and 4 fundamental hypermultiplets
arises as the world-volume theory on $N$ D3-branes
(extending along 0123)
in a Type IIB brane configuration 
in the space-time $\IR^{1,3}_{0123}
\times \IR^{2}_{45}/{\IZ_2^{\rm orient}}
\times \IC_{6789}^{2}/\IZ_2^{\rm orb}$. 
The orientifold group is
\be
G_{\rm orient} = 
\IZ_{2}^{\rm orb} + \IZ_{2}^{\rm orb}\Omega'=
\IZ_2^{\rm orb} \times \IZ_2^{\rm orient},
\la{orientiminusminus2}
\ee
where $\IZ_2^{\rm orb}=\{\identity,R_{6789}\}$ 
and $\IZ_2^{\rm orient}=\{\identity,R_{45}\Omega (-1)^{F_L}\}$.
$\IZ_2^{\rm orient}$ fixes the hyperplane $w=0$ 
(which corresponds to the classical position of the 
O7-plane and the 8 D7-branes, extending along 01236789), 
while $\IZ_2^{\rm orb}$ fixes the six-dimensional hyperplane $z_1=z_2=0$ 
(which corresponds to the classical position of the NS 5-branes 
extending along 012345 in the T-dual IIA configuration).
This fixed-point set of $G_{\rm orient}$ 
becomes $S^1\cup S^3 \subset S^5$ in the near-horizon limit. 

In the near horizon-limit 
the space-time changes to
$AdS_5\times S^5/(\IZ_2^{\rm orient}\times \IZ_2^{\rm orb})$, 
with the metric of the compact space given by
\be
&&d\widetilde{
\Omega}_5{}^2 = d\theta_1{}^2 + 
\cos^2\theta_1 \,\left(d\theta_2{}^2 +
\cos^2\theta_2\,d\phi_1{}^2+\sin^2\theta_2\,d\phi_2{}^2\,\right)+
\sin^2\theta_1\,
d\phi_3{}^2,\ret &&\qquad\qquad\theta_{1,2}\in
[0,\pi],\qquad\qquad
\phi_{1,2,3}\sim \phi_{1,2,3}+\pi, 
\la{metric2anti}
\ee
which is the angular part of the (flat) orbifold metric 
\be
ds^2= |dw|^2 + |dz_1|^2 + |dz_2|^2
\ee
with the orbifold actions 
$w\stackrel{\IZ_2^{\rm orient}}{\longrightarrow}-w$ 
and
$(z_1,z_2)\stackrel{\IZ_2^{\rm orb}}{\longrightarrow}(-z_1, -z_2)$. 
One way to derive this metric is to observe that 
the T-dual of the two NS 5-branes is a 
two-center Taub-NUT space, which in the limit where the
centers coincide becomes the orbifold  
$\IC_{6789}^{2}/\IZ_2^{\rm orb}$ 
(see Ref.~\cite{GremmKapustin} for a related discussion).

The rotation group  of the six-dimensional space
$\IR^{2}_{45}/ {\IZ_2^{\rm orient}}
\times \IC_{6789}^{2}/\IZ^{\rm orb}_2$ 
transverse to the D3-branes  is 
$\left[ {\rm SU}(2)_L/\IZ_2^{\rm orb} \right]
\times {\rm SU}(2)_R \times 
\left[ {\rm U}(1)_R/\IZ_2^{\rm orient} \right] \subset{\rm SO}(6)$. 
This is the only part of the full global symmetry group  
of the worldvolume theory,
\be
\underbrace{\su{2}_R\times\u{1}_R}_{R-{\rm
symmetry}}\times\underbrace{\su{2}_L\times\widetilde{\u{1}}_L}_{\rm
flavor~ sym.~of \,2 \, \Yasymm}
\times\underbrace{\su{4}_F\times\u{1}_B}_{\rm flavor~sym.~of \, 4\,\fund},
\la{glsymm}
\ee
that is manifest in the geometric description of the brane configuration.  
The $\su{4}_F\times\u{1}_B$ factor emerges as  the
gauge group on the D7-branes, 
while $\widetilde{\u{1}}_L$ emerges from the
low-energy dynamics of the twisted sector of $\IZ^{\rm orb}_2$. 

The supergravity
spectrum consists of three sectors: 
(a) the bulk sector, 
consisting of $AdS_5\times S^5$ supergravity states
invariant under 
$\IZ_2^{\rm orb}\times\IZ_2^{\rm orient}$,  
(b) the  $\IZ_2^{\rm orb}$ twisted sector, 
supported on $AdS_5\times S^1$, 
consisting of the KK reduction on $S^1$ 
of the six-dimensional $(2,0)$ tensor multiplet,
modded out by $\IZ_2^{\rm orient}$,
and (c) the D7-brane sector, 
consisting of the world-volume theory of 8 D7-branes
with topology $AdS_5\times S^3$, 
modded out by $\IZ_2^{\rm orb}$.

\vs{.1in}
\noindent{\large\bf (a) Bulk sector:}
\vs{.1in}

\noindent The bulk spectrum is obtained from 
the IIB spectrum on $AdS_5 \times S^5$ 
(Tables {\tablescalars} and \tablevectors)
by projecting onto states invariant under 
$\IZ_2^{\rm orb}\times\IZ_2^{\rm orient}$.
The orientifold breaks 
$\so{6}\simeq\su{4}_R
\to
\left[ \su{2}_L/\IZ_2^{\rm orb} \right] \times 
\su{2}_R \times \left[ \u{1}_R/\IZ_2^{\rm orient}\right]$. 
$\IZ_2^{\rm orb}$ coincides with the center of  ${\rm SU}(2)_L$, 
so even-dimensional representations are odd under $\IZ_2^{\rm orb}$, 
while odd-dimensional representations are even. 
$\IZ_2^{\rm orient}$ acts on fields with $\u{1}_R$ charge
$q_R$ as $e^{i\pi q_R/2}$, 
with an additional minus sign for states 
arising from ten-dimensional two-form fields. 
States must be invariant under both of these actions.

\noindent (1) {\it Scalar modes: }

\noindent Consider, for example, the supergravity scalar mode with $m^2 = -4$
in the ${\bf 20'}$ of $\su{4}_R$,
which couples to a $\Delta=2$ operator on the boundary. 
Under the  branching  
${\rm SO}(6) \rightarrow {\rm SU}(2)_L \times\su{2}_R \times \u{1}_R$, 
this representation decomposes as (see Table {\tblbranching})
\be
\rep{20'} \rightarrow
(3,3)_0+(2,2)_2+(2,2)_{-2}+(1,1)_4+(1,1)_{-4}+(1,1)_0.
\ee
$\IZ_2^{\rm orb}$ projects out all the states in even-dimensional
representations of $\su{2}_L$, 
while only states with $q_R=0$ mod 4 
survive the $\IZ_2^{\rm orient}$ projection.  This leaves
\be
(3,3)_0+(1,1)_4+(1,1)_{-4}+(1,1)_0.
\ee
Of these, only  $(1,1)_4$ and $(3,3)_0$ satisfy the constraint 
$\Delta = \vert q_R\vert/2 + d_R -1$ 
and thus correspond to CFT chiral primaries with $\Delta=2$. 
(The state $(1,1)_{-4}$, which also satisfies the above constraint,
is the complex conjugate of $(1,1)_4$, 
and is thus an {\it antichiral} primary field.)

There are two modes with $m^2 = -3$, 
in the ${\bf 10}$ and the ${\bf 50}$ of $\su{4}_R$, 
which couple to $\Delta=3$ operators on the boundary.
The ${\bf 50}$ decomposes into 
\be
\rep{50} \rightarrow
(4,4)_0+(3,3)_{\pm 2}+(2,2)_{\pm 4}+(2,2)_0+(1,1)_{\pm 6}+(1,1)_{\pm 2}.
\ee
Again, 
$\IZ_2^{\rm orb}$ removes the even-dimensional representations of 
$\su{2}_L$, while 
$\IZ_2^{\rm orient}$ only keeps $q_R=0$ mod 4, 
so no states survive the projection.
The ${\bf 10}$ decomposes into 
\be
\rep{10} \rightarrow (1,3)_2 +(3,1)_{-2} +(2,2)_0.
\ee
Now we have to take into account the 
extra minus sign from the action of $\Omega(-1)^{F_L}$ on
the ten-dimensional fields. 
$\IZ_2^{\rm orb}$ removes $(2,2)_0$, while $\IZ_2^{\rm orient}$ keeps only
those states with $q_R=2$ mod 4. 
This leaves $(1,3)_2+(3,1)_{-2}$, of which
only $(1,3)_2$ is chiral primary,  with $\Delta = 3$. 

Proceeding 
similarly
for the $\Delta = 4$ states, 
one finds the results summarized in Table {\tblsutwoanti}.

\noindent (2) {\it Vector modes: }
 
\noindent This set of modes is relevant because 
the corresponding operators couple to the conserved 
$\su{2}_L \times\su{2}_R \times \u{1}_R$ 
currents of the boundary CFT. 
The massless vector mode in the {\bf 15} of ${\rm SO}(6)$ 
(in Table \tablevectors) 
couples to the $\Delta=3$  $\su{4}_R$ current in the $\C{N}=4$
boundary theory. 
It decomposes into 
\be
\rep{15} \rightarrow (1,1)_0+(2,2)_2+(2,2)_{-2}+(3,1)_0+(1,3)_0, 
\ee
and, after the $G_{\rm orient}$ projection, 
we end up with three currents
$(1,1)_0,~(1,3)_0,~ (3,1)_0$, none of which is a chiral primary;
they are actually descendants of scalar chiral primaries. 
Note that $(1,1)_0+(1,3)_0$ have the correct quantum numbers to couple to the
$\su{2}_R\times\u{1}_R$ currents, while $(3,1)_0$ couples to the $\su{2}_L$
current. 

\vs{.1in}
\noindent {\large\bf (b) $\IZ_2^{\rm orb}$ twisted sector:}
\vs{.1in}

\noindent 
The twisted sector of $\IZ_2^{\rm orb}$ gives rise, 
in the near-horizon limit, to a set of states supported on
$AdS_5\times S^1$, 
where $S^1\subset S^5$ is left fixed by the $\IZ_2^{\rm orb}$ action.  
Gukov and Kapustin \cite{GukovKapustin},
using previous results in Ref.~\cite{Gukov}, 
have studied the corresponding sector of the
cousin $\sp{2N}\times\sp{2N}$ theory (see Sect. \ref{spsp} below),  
so we will follow their analysis closely.

Recall that the fixed-point set of $\IZ_2^{\rm orb}$, which in the
near-horizon limit looks like $AdS_5\times S^1$, 
corresponds in the Type IIB description 
to the classical position of the NS 5-branes in the T-dual configuration. 
Accordingly,
the resulting supergravity spectrum is obtained by the KK reduction on
$AdS_5\times S^1$ of the six-dimensional $(2,0)$ tensor multiplet, 
the low-energy world-volume theory of a {\it single} NS 5-brane. 
In addition, one has to project the resulting states by
$\IZ_2^{\rm orient}$. 

Let us briefly review, following the analysis in \cite{DouglasMoore}, how
this comes about. Start with Type IIB string theory on the orbifold
$\IC^2_{6789}/\IZ_2^{\rm orb}$ \footnote{\baselineskip12pt 
D3-branes probing this background 
give rise to a finite $\C{N}=2$ $\su{N}\times\su{N}$ theory
\cite{KachruSilver,AlbionNikitaCumrun}, whose dual supergravity
description has been studied in Ref.~\cite{OzTerning,Gukov}. 
The IIB configuration itself is T-dual to 
the M-theory construction described in Ref.~\cite{WittenMtheory}.}. 
The massless untwisted sector contains the
$(2,0)$ gravity multiplet and two tensor multiplets. 
There is in addition
a third tensor multiplet supported on the six-dimensional hyperplane fixed
by the orbifold action. 
The orbifold breaks the ten-dimensional transverse
Lorentz group $\so{8}$ to 
$[\su{2} \times \su{2}]_{\rm little~group} \times \su{2}_L\times\su{2}_R$, 
where 
$[\su{2} \times \su{2}]_{\rm little~group}$ 
is the little group in six dimensions and
$\su{2}_L\times\su{2}_R\simeq \so{4}$ is the rotation group in
$\IC^2_{6789}$. In terms of these groups,  the tensor
multiplet in the twisted sector contains the (bosonic) fields 
listed in Table 8.

\begin{center}
\begin{tabular}{||c|c|c|c||}
\hline\hline
Sector & $[\su{2} \times \su{2}]_{\rm little~group}$
& $\su{2}_L\times\su{2}_R$ & 10d origin 
\\
\hline
NS-NS & (1,1) & (1,1) & NS-NS 2-form 
\\
\hline
NS-NS & (1,1) & (1,3) & graviton
\\
\hline
R-R & (3,1) & (1,1) & 4-form
\\
\hline
R-R & (1,1) & (1,1) & R-R 2-form 
\\
\hline\hline
\end{tabular}
\end{center}
\centerline{\small{\bf Table {\tbltensor}:} 
$(2,0)$ tensor multiplet}
\vspace{0.5cm}
 
Now orientifold the above IIB setup by $\Omega$
only to obtain Type I on 
$\IC^2_{6789}/\IZ_2^{\rm orb}$. The resulting Type I configuration
contains an O9-plane, D9-branes (to cancel the O9 charge), and a
six-dimensional plane where the $\Omega$ projection of the original IIB
twisted sector propagates. This six-dimensional theory has $(1,0)$
supersymmetry, which corresponds to
$\C{N}=2$ in four dimensions. If we now add probe D5-branes without
breaking any further supersymmetry, the resulting configuration is
related by double T-duality along both $x^4$ and $x^5$ to the IIB
configurations,
leading to $\su{N}+2\Yasymm+4\,\fund$ and $\sp{2N} \times\sp{2N}$. 

There are two inequivalent ways to perform the $\Omega$
projection in the above setup. In the first one, leading to theories with
vector structure, $\Omega$ acts with an additional minus sign on the
twisted sector of $\IZ_2^{\rm orb}$. The $\Omega$ projection breaks the
$(2,0)$ tensor multiplet into a $(1,0)$ tensor multiplet and a $(1,0)$
hypermultiplet, and projects out the hypermultiplet. The
remaining $(1,0)$ tensor multiplet corresponds to the twisted sector of the
$\sp{2N}\times\sp{2N}$  theory in the double T-dual IIB
configuration. 
In the second possibility $\Omega$ acts without the extra minus sign (and
thus projects out the 
$(1,0)$ tensor multiplet, keeping the hypermultiplet), leading to a model
without vector structure which is T-dual to the $\su{N}+2\,\Yasymm+4\,\fund$
theory.

Going back to our IIB setup, we need only perform the KK reduction of the
$(2,0)$ tensor multiplet on $S^1$
and then project by  $\IZ_2^{\rm orient}$. The reduction on $S^1$
gives an additional $\u{1}_R$ symmetry 
(coming from rotations of the circle), 
with a state of KK momentum $\ell \in \IZ$ 
carrying $\u{1}_R$ charge $q_R=2\ell$ \cite{Gukov,GukovKapustin}.
Reducing the $(2,0)$ tensor multiplet in Table {\tbltensor} on $S^1$,
we obtain the following states \cite{Gukov,GukovKapustin}: 

\noindent (1) {\it  Scalar modes:}

\noindent The triplet of scalars  $(1,1;1,3)$ of 
$[\su{2}\times\su{2}]_{\rm little~ group}\times\su{2}_L\times\su{2}_R$ 
gives rise to a tower of
$\su{2}_R$ triplets $(1,3)_{2\ell}$, where
$\ell \geq 0$, with masses
$m^2\,=\,\ell^2-4$, coupling to chiral operators with 
$\Delta=\ell+2$ and
$R=2\ell$ in the CFT.  
(States with $\ell<0$ couple to antichiral operators on the boundary.) 
$\IZ_2^{\rm orient}$ acts as $e^{i\pi q_R/ 2}\,=\,e^{i\pi \ell}$, 
so only states with $\ell=0$ mod 2 or $q_R=0$ mod 4
survive the $\IZ_2^{\rm orient}$ projection, 
leaving a tower of states
\be
&&\Delta= 2\qquad\qquad(1,3)_0, \quad m^2\,=\,-4,\nn\\
&&\qquad\, 4\qquad\qquad(1,3)_4,\quad m^2\,=\,0,\ret 
&&\qquad\,\,\vdots
\ee
These obey the constraint  \eqs{chiralprimaries}
and so correspond to chiral primary operators.

The pair of scalars $(1,1;1,1)$ and $(1,1;1,1)$ coming from the
ten-dimensional  2-forms 
give rise to two families of KK states with
quantum numbers $(1,1)_{2\ell}$: 

(i) $m^2\,=\,\ell^2-4\ell\,,\,\Delta=\ell\,\,,\,\,q_R=2\ell,\,\,\ell\geq 2$,

(ii) $m^2\,=\,\ell^2+4\ell,,\,\Delta=\ell+4\,\,,\,\,q_R=2\ell,\,\,\ell\geq 0$.

\noindent All the states in family (i) satisfy the constraint 
\eqs{chiralprimaries} and are thus potentially chiral primaries, 
while none of the states in family (ii) satisfy the constraint. 
Including the extra minus sign coming from $\Omega(-1)^{F_L}$, 
$\IZ_2^{\rm orient}$ acts on the states in family (i) as
$e^{i\pi\left(q_R/2 +1\right)}\,=\,e^{i\pi (\ell+1)}$, 
so the projection keeps only those states with  
$q_R=2\mod4$ or $\ell=1\mod2$, leaving the tower of states 
\be
&&\Delta=3\qquad (1,1)_6,\quad m^2=-3, \nn\\
&&\qquad\,5\qquad(1,1)_{10},\quad m^2=5.\ret
&&\qquad\,\,\vdots
\ee

\noindent (2) {\it Vector modes:}

\noindent By naive dimensional reduction on $S^1$ 
we would expect to obtain a vector state from the 2-form
$(3,1;1,1)$. This is indeed the case \cite{GukovKapustin}, but only for
zero KK momentum ($\ell=0$). The 
resulting state is a massless vector with quantum numbers 
$(1,1)_0$ under $\su{2}_L\times\su{2}_R\times\u{1}_R$,
which survives the $\IZ_2^{\rm orient}$ projection. 
This massless vector state couples to a $\Delta=3$ 
conserved current on the boundary, 
which we can identify with the 
$\widetilde{\u{1}}_L$ current that was missing from
the bulk sector.

\np
\noindent {\large\bf (c) D7-brane sector:}
\vs{.1in}

\noindent The AdS compactifications of Type IIB string theory T-dual to 
Type IIA O$6^{-}$--\,O$6^{-}$ configurations include, as described
above, 8 D7-branes with world-volume $AdS_5\times S^3$, where  
$S^3\subset S^5/G_{\rm orient}$. The low-energy excitations on these 
7-branes comprise a $7+1$ dimensional $\C{N}=1$ vector multiplet
with a gauge group $\C{G}_7\subset \u{8}$ that depends on the details 
of $G_{\rm orient}$.  As explained in Ref.~\cite{Ansar2}, 
the supergravity spectrum includes the KK modes of this vector multiplet 
on $AdS_5\times S^3$, which are dual to primary operators of 
the CFT charged under the global\footnote{\baselineskip12pt \noindent 
{}From the point of view of the
boundary field theory, the gauge group on the 7-branes corresponds to the
{\it global} flavor symmetry group of the hypermultiplets in the fundamental
representation of the gauge group on the D3-branes.} symmetry group
$\C{G}_7$.  

The 7-brane spectrum for the $\sp{2N}$ + $\Yasymm$ + $4\,\fund$ theory
(which has $\C{G}_7=\so{8}$) has been computed in Ref.~\cite{Ansar2},
and for the $\sp{2N}\times\sp{2N}$ theory 
(which has $\C{G}_7=\so{4}\times\so{4}$) in Ref.~\cite{GukovKapustin}.
In the latter case, as in the $\su{N}$ + $2\Yasymm$ + $4\,\fund$ theory 
that we are now considering (which has $\C{G}_7=\u{4}$), 
the spectrum is the $\IZ_{2}^{\rm orb}$ 
projection of the $\C{G}_7=\so{8}$ case,
which itself is the $\IZ_{2}^{\rm orient}$ projection of the
$\C{G}_7=\u{8}$ case (corresponding to 8 7-branes in a flat background). 
This comes about as follows.  Start with a collection of 8 D7-branes in 
$\IR^{1,9}$. As explained above, the 77 open string sector gives rise to
a 7+1 dimensional $\C{N}=1$ SYM theory with gauge group $\C{G}_7=\u{8}$. 
Now add an orientifold 7-plane parallel to the 7-branes, which induces a 
$\IZ_{2}^{\rm orient}$ projection on the 7-brane worldvolume theory that
is best described in terms of the action of the corresponding projection 
matrix $\gamma_{\Omega' 7}$ on the Chan-Paton factors of the 77 strings.
Gimon and Polchinski \cite{GimonPolchinski} have found 
(in the T-dual Type I setup) 
\be
\gamma_{\Omega' 7}=\identity_8,
\la{project1}
\ee
which enforces the projection $\u{8}\to\so{8}$  
(see, for example, \cite{DabolkarRev}). 
To see this in more detail, let $M$ 
be a hermitian $8\times 8$  Chan-Paton matrix. 
$\IZ_2^{\rm orient}$ enforces the projection 
\be
M=-\gamma_{\Omega' 7} ^{T}M^T \gamma_{\Omega' 7}^{-1} =-M^T,  
\ee
from which it follows that $M$ is an antisymmetric $8 \times 8$ matrix which
can be parametrized as
\be
M\,=\, \pmatrix{
 {\sss{a}}_1& \sss{b}\cr -\sss{b}^T
& \sss{a}_2\cr},
\la{eme}
\ee
where $\sss{a}_1, \sss{a}_2$ are $4 \times 4$ antisymmetric matrices
and $\sss{b}$ is a $4 \times 4$ matrix, giving a total of 
$6+6+16=28$ independent real coefficients, 
as appropriate for ${\rm SO}(8)$.

Next consider the $\IZ_2^{\rm orb}$ projection. 
There are two different ways of performing this projection 
\cite{GimonPolchinski}, 
one breaking $\so{8}\to\so{4}\times\so{4}$ (and thus corresponding to the
$\sp{2N}\times\sp{2N}$ theory), 
the other breaking $\so{8}\to\u{4}$ 
(and giving $\su{N}$ + 2$\Yasymm$ + 4$\,\fund$\,). There are 
correspondingly two different projection matrices, 
\be
\gamma_{\theta 7}^{\rm Sp\times Sp}= \Biggr(\begin{tabular}{cc}
\hskip-2pt $\identity_4$ &
\hskip-8pt 0
\\[-.05in]
\hskip-7pt 0 & \hskip-10pt $-\identity_4$
\end{tabular}
\hskip-5pt\Biggr),\qquad{\rm and}\qquad 
\gamma_{\theta 7}^{\rm SU}= \Biggr(\begin{tabular}{cc}
\hskip-5pt 0& \hskip-8pt$i\identity_4$\\[-.05in]\hskip-5pt$-i\identity_4$
&\hskip-8pt 0
\end{tabular}
\hskip-5pt\Biggr).
\la{project2}
\ee

In the present case the appropriate matrix is the second one, 
which enforces on $M$ \eqs{eme} the projection 
\be
M= \gamma_{\theta 7}^{\rm SU}\, M \, (\gamma_{\theta 7}^{\rm SU})^{-1} 
\,\Rightarrow \, M= M_{\u{4}} =  \pmatrix{
\sss{a}& \sss{s}\cr -\sss{s}
& \sss{a}\cr}, 
\ee
where $\sss{a}$ is a $4\times 4$ antisymmetric
matrix,  and \sss{s} is a $4\times 4$ symmetric matrix, giving a total of
$6+10=16$ real coefficients, as appropriate for $\u{4} \sim
\su{4}_F \times \u{1}_B$, the correct flavor group for 4
hypermultiplets in the fundamental representation of $\su{N}$ (with
$N\geq3$). 

Under the branching 
${\rm SO}(8) \rightarrow {\rm SU}(4)_F \times\u{1}_B$ 
the adjoint representation of $\so{8}$ decomposes as follows 
\be
\rep{28} \rightarrow 1_0\,+\,15_0\,+\,6_2\,+\,6_{-2},
\ee
where $1_0\,+\,15_0$ is the adjoint of $\su{4}_F\times\u{1}_B$. 
In terms of $M$ \eqs{eme} in the $\rep{28}$ of ${\rm SO}(8)$, 
this comes about as follows. 
Let $M=M_{\u{4}}+\widetilde{M}$, which defines $\widetilde{M}$
as a $8\times8$ matrix in the $6_2\oplus6_{-2}$ of 
${\rm SU}(4)_F\times\u{1}_B$.
%
Then 
\be
\widetilde{M} =  \pmatrix{
\widetilde{\sss{a}} & \sss{a}_b \cr -\sss{a}_b^T
& -\widetilde{\sss{a}}\cr},\qquad\quad {\rm with}\qquad
\begin{tabular}{l}
$\sss{a}_b=\sss{b}-\sss{s}$, \\[-.1in] 
and
\\[-.1in]  
$\widetilde{\sss{a}}=(\sss{a}_1 -\sss{a}_2)/2$, \end{tabular}
\ee
from which it is easy to see that $\Big[\gamma_{\theta 7}^{\rm SU},
M_{\u{4}}\Big]\,=\,0$ and  
$\Big\{\gamma_{\theta 7}^{\rm SU}, \widetilde{M}\Big\}\,=\,0$.

To work out the D7-brane excitations we start with the $d=8$, $\C{N}=1$
vector multiplet in the adjoint of ${\rm SO}(8)$, 
containing a vector $A$, a complex scalar  $z$, and fermions.  
The field $z$ is in the  $(1,1)_2$ of\, ${\rm SU}(2)_L \times {\rm SU}(2)_R
\times \u{1}_R$, where $\u{1}_R$, the rotation group of the
two-dimensional space transverse to the 7-branes, is the $R$-symmetry group 
in $d=8$, and ${\rm SU}(2)_L \times {\rm SU}(2)_R\simeq \so{4}$ is the
isometry group of $S^3$.  Now reduce  this multiplet on $AdS_5 \times S^3$,
to obtain \cite{Ansar2} a KK tower of $d=5$ vector multiplets. The scalar
$z$ gives rise to a tower of  scalar modes in representations $(k,k)_2$ of 
${\rm SU}(2)_L \times {\rm SU}(2)_R\times \u{1}_R$ for
$k=1,2,3,\ldots$. The
$d=8$ vector field $A$ decomposes into $AdS_5$ vectors with ${\rm SU}(2)_L
\times {\rm SU}(2)_R\times \u{1}_R$ quantum numbers $(k,k)_0$, and two
families of real scalar fields
$(k,k+2)_0\oplus(k+2,k)_0$. All these states are in the adjoint
representation (\rep{28}) of $\so{8}$. As explained in
Ref.~\cite{Ansar2}, they are all in short multiplets of the
superconformal algebra,  with their masses uniquely determined 
in terms of their ${\rm SU}(2)_R \times\u{1}_R$ representations.
The lowest component of each such short multiplet, 
which is a real scalar field $(k,k+2)_0$ with $\Delta=k+1$,  
is automatically a chiral primary state of the boundary CFT. 

The other components of a given short multiplet 
(which are protected, even though they are not chiral primaries), 
arise by acting with supercharges on the lowest component. 
The first three such multiplets (containing chiral primaries with
$\Delta\leq4$) are summarized 
in the following table:
\vglue.1in
\vbox{
\begin{center}
\begin{tabular}{cccccc}
{} & k & 1 & 2 & 3 & $\cdots$ 
\\
$\Delta$ & {} & {} & {} & {} & {}
\\
2 & {} & \fbox{$(1,3)_0^{\rm scalar}\atop m^2 = -4$}
&
\rule{10mm}{.1mm} &
\rule{10mm}{.1mm} & {}
\\ [.1in]
3 & {} & $(1,1)_2^{\rm scalar}\oplus (1,1)_0^{\rm vector}\atop
m^2=-3\quad\,\,\,\, m^2=0$ &
\fbox{$(2,4)_0^{\rm scalar}\atop m^2=-3$} &  \rule{10mm}{.1mm} & {}
\\ [.1in]
4 & {} & \rule{10mm}{.1mm} &  $(2,2)_2^{\rm scalar}\oplus
(2,2)_0^{\rm vector}\atop m^2=0\qquad m^2=3$ & \fbox{$(3,5)_0^{\rm
scalar}\atop m^2 =0$} & {} 
\\ [.1in]
5 & {} & \rule{10mm}{.1mm} & \rule{10mm}{.1mm} & $(3,3)_2^{\rm
scalar}\oplus (3,3)_0^{\rm vector}\atop m^2=5\qquad m^2=8$ & {}
\\ [.1in]
6 & {} & \rule{10mm}{.1mm} & \rule{10mm}{.1mm} & $(3,1)_0^{\rm scalar}\atop
m^2 =15$ & {}
\\
$\vdots$ & {} & {} & {} & {} & {}
\end{tabular}
\end{center}
\begin{center}
\begin{minipage}{4.7in}
\baselineskip12pt
{\noindent\small \fbox{Boxed} states are chiral primaries. The massless
vector $(1,1)_0$ in the $k=1$ family couples to the 
${\rm SO}(8)$ current of the boundary CFT.}
\end{minipage}
\end{center}
}
 
We now have to project by $\IZ_2^{\rm orb}$, which acts 
on the above states according to their ${\rm SU}(2)_L$ representation
(such that odd-dimensional representation are even, while
even-dimensional representation are odd) and also as a conjugation
by $\gamma_{\theta 7}^{\rm SU}$. As explained above, this breaks 
${\rm SO}(8)\to\u{4}\simeq {\rm SU}(4)_F\times\u{1}_B$ 
and 
$\rep{28} \rightarrow 1_0+15_0+6_2+6_{-2}$. 
The states in the $1_0+15_0$ representation are even
under conjugation by
$\gamma_{\theta 7}^{\rm SU}$ (their Chan-Paton matrix
$M_{\u{4}}$ commutes with 
$\gamma_{\theta 7}^{\rm SU}$),  while the states in representations
$6_2+6_{-2}$ are odd. 
Thus  
$\IZ_2^{\rm orb}$ keeps states in odd-dimensional
representations $1,3,5,\ldots$ of
$\su{2}_L$ that transform in the adjoint $1_0+15_0$ of
$\su{4}_F\times\u{1}_B$, 
and states in even-dimensional representations 
$2,4,\cdots$ of ${\rm SU}(2)_L$ that transform in  the $6_2+6_{-2}$ of
${\rm SU}(4)_F\times\u{1}_B$. 

\vbox{
\begin{center}
\begin{tabular}{||c|c|c|c|c|c|c||}
\hline\hline
$\Delta$ &Sector & $\su{4}_R$ & 
$(d_L, d_R)_{q_R}$ & 
$\su{4}_F \times \u{1}_B$ & $m^2$ &  CFT operator  
\\
\hline\hline
2 & bulk & $\rep{20'}$ & $(1,1)_4$ &\rule{5mm}{.1mm}& $-4$ &
				 ${\rm tr} \,\Phi^2$ \\ 
2 & bulk & $\rep{20'}$ & $(3,3)_0$ &\rule{5mm}{.1mm}& $-4$ & 
			$\tilde A_{(i} A_{j)}$ \\ 
2 & twisted & &  $(1,3)_0$ &\rule{5mm}{.1mm}& $-4$ & $\tilde A_{[i} A_{j]}$ 
\\ 
2 & D7 & & $(1,3)_0$ & $1_0\oplus 15_0$ & $-4$ & $\tilde Q^I Q_J$ 
\\ 
3 & bulk & \rep{10} & $(1,3)_2$ &\rule{5mm}{.1mm}& $-3$ & ${\rm tr}\,\lambda\lambda$ 
\\ 
3 & twisted & & $(1,1)_6$ &\rule{5mm}{.1mm}& $-3$ & ${\rm tr}\, \Phi^3$   
\\ 
3 & D7 & & $(2,4)_0$ & $6_2\oplus 6_{-2}$ & $-3$ &
$Q_{[I} \tilde A_i Q_{J]} \oplus \tilde Q^{[I} A_i\tilde Q^{J]}$ 
\\  
4 & bulk & \rep{105} & $(1,1)_8$ &\rule{5mm}{.1mm}& 0 & ${\rm tr}\,\Phi^4 $ 
\\ 
4 & bulk & \rep{105} & $(3,3)_4$ &\rule{5mm}{.1mm}& 0 & $\tilde  A_{(i} \Phi^2 A_{j)}$
\\ 
4 & bulk & \rep{105} & $(5,5)_0$ &\rule{5mm}{.1mm}& 0 & $\tilde  A_{(i} A_j \tilde A_k
A_{l)}$
\\
4 & twisted & & $(1,3)_4$ &\rule{5mm}{.1mm}& 0 & $\tilde A_{[i} \Phi^2 A_{j]}$ 
\\ 
4 & D7 & & $(3,5)_0$ & $1_0\oplus 15_0$ & 0 & 
$\tilde Q^I A_{(i} \tilde A_{j)} Q_J$  
\\
\hline\hline
\end{tabular}
\end{center}
\begin{center}
\begin{minipage}{5.8in}
\baselineskip12pt
{\noindent\small{\bf Table {\tblsutwoanti}}:
Scalars with $\Delta \le 4$ in chiral primary representations of
$\su{N}$\,\, w/\,\,\,$2\,\Yasymm\oplus4\,\fund\,$.\\
$d_R$, $d_L$ denote the dimensions of the
$\su{2}_R$, $\su{2}_L$ representations; 
$q_R$ is the $\u{1}_R$ charge;
$i,j$ are $\su{2}_L$ indices, and 
$I,J$ are $\su{4}_F$ indices.}
\end{minipage}
\end{center}
}

The chiral primaries operators with $\Delta \le 4$ 
from all the sectors are compiled in Table {\tblsutwoanti}; 
anti-chiral primary operators are also present in the spectrum 
(and protected), and are complex conjugates of the chiral primaries.
The last column of Table {\tblsutwoanti}
lists the CFT operator that correspond to each supergravity state.\footnote{
\baselineskip14pt
The identification of the operator corresponding to the $(1,3)_4$ state
is not completely unambiguous, 
as tr $\lambda \lambda \Phi$ has the same quantum numbers.}
(As before, the highest component of each $\su{2}_R$ representation is listed.)
The perturbative field theory description of the worldvolume 
$\su{N}+2\,\Yasymm + 4\,\fund$ theory
includes an $\C{N}=2$ vector multiplet in the adjoint representation
and two antisymmetric and four fundamental $\C{N}=2$ hypermultiplets.
As described in the previous section,
each of these $\C{N}=2$ multiplets consists of   
several $\su{2}_R$ representations, 
with definite scaling dimension $\Delta$.
Those that comprise the chiral primary operators 
in Table {\tblsutwoanti} are listed in
Table {\tblsutwoantifield} below.

\vbox{
\begin{center}
\begin{tabular}{||l|c|c|c|c|c||}
\hline\hline
$\Delta$ & $\su{N}$ & $\su{2}_L \times \su{2}_R \times \u{1}_R $ & 
$\widetilde{\u{1}}_L$  &
$\su{4}_F\times \u{1}_B$ & CFT field  
\\
\hline\hline
${3\over 2}$ 	&  adjoint 	& $(1,2)_1$ & 0 & $1_0$  & $\lambda$ \\ 
{1} 		& adjoint 	& $(1,1)_2$ & 0 & $1_0$  & $\Phi$ \\  
{1} 		& $\Yasymm$ 	& $(2,2)_0$ & 1 & $1_0$  & $A_i$ \\  
{1} 		& $\fund$  	& $(1,2)_0$ & 0 & $4_1$  & $Q_I$ \\  
\hline\hline
\end{tabular}
\end{center}
\begin{center}
\baselineskip12pt
{\noindent\small{\bf Table {\tblsutwoantifield} }: 
$\su{2}_R$ multiplets in perturbative
$\su{N}$\,\, w/\,\,\,$2\,\Yasymm\oplus4\fund,$}
\end{center}
}
\vspace{0.5cm}

Before examining the table of chiral primaries further,
let us consider how the flavor symmetry group 
$\su{4}_F\times\u{1}_B\times\su{2}_L\times\widetilde{\u{1}}_L$  
acts on the hypermultiplet fields.
The $\su{4}_F\times\u{1}_B$ factor 
affects only the fundamental hypermultiplets $Q_I$ 
(and $\tilde Q^I$ in the complex conjugate representation),
$I=1,\ldots,4$,
with $\su{4}_F$ rotating the $I$ index,
so that $Q_I$ ($\tilde Q^I$) transforms in the $4$ ($\bar 4$), 
and with $\u{1}_B$ acting as 
\be
Q_I\to e^{i\alpha}Q_I,\qquad\qquad\tilde Q^J\to e^{-i\alpha}\tilde Q^J,
\la{baryonnum}
\ee 
The $\su{2}_L\times\widetilde{\u{1}}_L$ factor acts only
on the two antisymmetric hypermultiplets,
each of which comprises a pair of $\C{N}=1$ chiral superfields
in the antisymmetric and the complex conjugate antisymmetric
representations respectively, which in obvious notation are 
\be
\{A_1,\tilde A^1\}\,\,,  \{A_2,\tilde A^2\}
\ee
Now $(A_1,A_2)$ form a $\u{2}_L\sim {\rm SU}(2)_L\times \widetilde{\u{1}}_L$
doublet, whereas 
$(A_1, (\tilde A^1)^\dagger)$ and   
$(A_2, (\tilde A^2)^\dagger)$ form ${\rm SU}(2)_R$ doublets. 
$\u{2}_L$ is a flavor symmetry
for the 2 antisymmetric hypermultiplets, 
as can be seen from the piece of the tree-level superpotential
\be
\C{W}\,=\,{\rm tr}\,(\tilde A^1 \Phi A_1\,+\,  \tilde A^2 \Phi A_2),
\ee
involving the antisymmetric hypermultiplets. 
(As we are considering the
semiclassical limit of the ${\cal N}=2$ theory, we need not worry about
the fate of the superpotential in the strong-coupling regime.)
${\rm SU}(2)_L$ arises as a geometric symmetry in the $AdS$ picture,
while $\widetilde{\u{1}}_L$ corresponds to the abelian massless vector we
found in the twisted sector of the $\IZ_2^{\rm orb}$. 

We can give tree-level mass terms to the 2 antisymmetric hypermultiplets as
follows: 
\be
m_1\tilde A^1A_1\,+\,m_2\tilde A^2A_2
\la{massterm}
\ee
If $m_1\not= m_2$, $\u{2}_L$ is broken to 
$\u{1}_L \times\widetilde{\u{1}}_L$, 
where each of the $\u{1}$'s acts as follows
\begin{equation}
\begin{tabular}{lllll}
\multicolumn{2}{c}{$\u{1}_L$} &{}&
\multicolumn{2}{c}{$\widetilde{\u{1}}_L$} \\
$A_1\rightarrow e^{i\alpha} A_1$ & $\tilde A^1\rightarrow
e^{-i\alpha}
\tilde A^1$ & {} &
$A_1\rightarrow e^{i\beta} A_1$ & $\tilde A^1\rightarrow e^{-i\beta}
\tilde A^1$
\\
$A_2\rightarrow e^{-i\alpha} A_2$ & $\tilde A^2 \rightarrow
e^{i\alpha}
\tilde A^2$ & {} &
$A_2 \rightarrow e^{i\beta} A_2$ & $\tilde A^2 \rightarrow e^{-i\beta}
\tilde A^2$
\end{tabular}
\la{uunos2anti}
\end{equation}
If $m_1=m_2$, the full $\u{2}_L$ is restored. 
To see this, define $\globalmass = (m_1-m_2)/2$ 
(known as the {\it global mass} in the M-theory description 
\cite{WittenMtheory,UrangaElliptic,Elliptic}) 
and $\mu=(m_1+m_2)/2$ so that \eqs{massterm} becomes
\be
&&\globalmass\,(\tilde A^1A_1-\tilde A^2A_2)\,+\,\mu\,(\tilde A^1A_1+\tilde
A^2A_2) \\
&=&\globalmass\,(\tilde A_2 A_1+\tilde A_1 A_2)\,
+\,\mu\,(\tilde A_2 A_1-\tilde A_1 A_2), 
\la{massdeform}
\ee
where in the last line, 
we have lowered the $\su{2}_L$ indices with the $\epsilon$ tensor.

Returning to Table {\tblsutwoanti},
we find two relevant and one marginal
deformations corresponding to the Coulomb branch moduli 
(the Casimirs \hbox{tr\,$\Phi^n$}, with $n\leq4$), 
and a relevant $\Delta=3$ deformation corresponding to the gaugino 
condensate. 
We also find three relevant $\Delta=2$ deformations 
containing the mass operators of the two
antisymmetric hypermultiplets and the mass operator of the
fundamental hypermultiplets. 
Comparing eq.~\eqs{massdeform} with Table {\tblsutwoanti},
we observe 
that the $(3,3)_0$ operator arising from the bulk supergravity sector
corresponds to a deformation of the global mass $\globalmass$,
whereas the $(1,3)_0$ operator arising from the twisted sector
corresponds to a deformation of the average mass $\mu$ of the 
antisymmetric hypermultiplets.
{}From field theory considerations, we should expect to  
find at least two  marginal deformations with zero $R$-charge 
corresponding to the change of the gauge coupling and theta angle. 
Actually, these deformations correspond to a CFT complex scalar operator
\hbox{tr\,($F^2 + iF*F$)}, which is a descendant (hence not a chiral
primary) of \hbox{tr\,$\Phi^2$}. 
On the supergravity side, this state is a singlet 
of $\su{4}_R$ with $\Delta=4$ and $m^2=0$, 
which survives the $G_{\rm orient}$ projection 
but is not a chiral primary.

Finally, Table
{\tblsutwoantivect} contains a summary of the massless vector states we
have found, which couple to the
$\Delta=3$ (conserved) currents of the full global symmetry group. 
Notice that a $(1,1)_0$   massless vector state in
the adjoint of ${\rm SU}(4)_F \times \u{1}_B$ 
in the D7-brane sector
survives the $\IZ_2^{\rm orb}$ projection, 
and couples to the 
${\rm SU}(4)_F \times \u{1}_B$ 
flavor current of the boundary CFT. 

\vbox{
\begin{center}
\begin{tabular}{||c|c|c|c|c|c|c||}
\hline\hline
$\Delta$ & Sector & $\su{4}_R$ & 
$({d_L},{d_R})_{q_R}$ 
& $\su{4}_F \times \u{1}_B$ & 
$m^2$ &  CFT current  
\\
\hline\hline
3 & bulk & \rep{15} & $(1,3)_0$ &\rule{5mm}{.1mm}& $0$ & $\su{2}_R$ 
\\ 
3 & bulk & \rep{15} & $(1,1)_0$ &\rule{5mm}{.1mm}& $0$ & $\u{1}_R$ 
\\ 
3 & bulk  & \rep{15} &  $(3,1)_0$ &\rule{5mm}{.1mm}& $0$ & $\su{2}_L$ 
\\ 
3 & twisted & & $(1,1)_0$ &\rule{5mm}{.1mm}& $0$ & $\widetilde{\u{1}}_L$ 
\\ 
3 & D7 & & $(1,1)_0$ & $1_0\oplus 15_0$ & $0$ & $\su{4}_F\times\u{1}_B$
\\
\hline\hline
\end{tabular}
\end{center}
\centerline{\small{{\bf Table {\tblsutwoantivect}}: 
Massless vector states for  
$\su{N}$\,\, w/\,\,\,$2\,\Yasymm\oplus4\,\fund\,$.}}
}

\subsection{$\sp{2N}\times\sp{2N}$ + 1 bifundamental + 4 fundamental
hypermultiplets}
\la{spsp}

The analysis of the $\sp{2N}_1\times\sp{2N}_2$  theory 
with hypermultiplet content 
$(\fund{}_{1},\fund{}_{2})\oplus2\,\fund\,{}_{1}\oplus 2\,\fund\,{}_{2}$
proceeds in parallel with the 
$\su{N} + 2 \Yasymm + 4 \fund$ theory,
as they both share the same IIB background 
(with a few differences, discussed in the previous section,
that serve to distinguish the two). 
Gukov and Kapustin \cite{GukovKapustin} have analyzed this
theory in detail, 
so we will just review their results 
and make a few remarks to complete and clarify them. 

The orientifold group and metric are the same as in Sect.~\ref{sect2anti}
(see Eqs. \eqs{orientiminusminus2}  and \eqs{metric2anti}).
As before, $\IZ_2^{\rm orb}$ fixes a six-dimensional hyperplane, 
while $\IZ_2^{\rm orient}$ fixes an eight-dimensional hyperplane. 
The fixed-point set becomes  $S^1\cup S^3 \subset S^5$ 
in the near-horizon limit. 
The global symmetry group, however, is different   
\be
\su{2}_R\times\u{1}_R\times\su{2}_L
\times\so{4}\times\so{4},
\la{glsymmspsp}
\ee
and reflects the differences between the two cases. 
The $\so{4}\times\so{4}$ factor, 
which emerges in the brane configuration as the
gauge group on the D7-branes, 
is the flavor symmetry of the fundamental
hypermultiplets, two for each $\sp{2N}$  factor
(the enhancement from $\su{2}$ to $\so{4}$ for
self-conjugate representations was described 
in Sect.~\ref{sectspanti}).
The $\su{2}_L$ factor 
is the flavor symmetry of the bifundamental hypermultiplet,
as explained below.
Note that the {\it geometric} part of the global symmetry,
namely, $\su{2}_L\times \su{2}_R\times\u{1}_R$,
is the same in both cases, 
a natural result given that both
theories have the same {\it geometric} 
description\footnote{\baselineskip14pt In Ref.~\cite{GukovKapustin}, 
only the Cartan subgroup
$\u{1}_L\subset\su{2}_L$ is considered, 
even though their tables contain all the states 
needed to form $\su{2}_L$ multiplets. 
We will explicitly use the full $\su{2}_L$ symmetry below, 
as this emphasizes the similarity with the
$\su{N}+2\,\Yasymm+4\,\fund$ case.}. 

As in the previous section, the supergravity
spectrum consists of three sectors:

\vs{.1in}
\noindent{\large\bf (a) Bulk sector:}
\vs{.1in}

\noindent This sector consists of $AdS_5\times S^5$ supergravity states
invariant under $\IZ_2^{\rm orb}\times\IZ_2^{\rm orient}$, 
and is identical to that in the $\su{N}+2\,\Yasymm+4\,\fund$ theory, 
although the field theory interpretation of
the resulting states obviously differs (see Table {\tblspxsp}).

\vs{.1in}
\noindent {\large\bf (b) $\IZ_2^{\rm orb}$ twisted sector:}
\vs{.1in}

\noindent 
The $\IZ_2^{\rm orb}$ twisted sector, 
supported on $AdS_5\times S^1$, 
consists of the KK reduction on $S^1$ of 
the six-dimensional $(2,0)$ tensor multiplet,
modded out by $\IZ_2^{\rm orient}$. 
This is identical to the previous section,
except that $\IZ_2^{\rm orient}$ acts on this sector 
with an additional minus sign. 

\noindent (1) {\it Scalar modes:}

\noindent Recall that the triplet of scalars  $(1,1;1,3)$ 
gave rise upon KK reduction on $S^1$ 
to a tower of $\su{2}_R$ triplets $(1,3)_{2\ell}$, 
$\ell \ge 0$, with $m^2 = \ell^2 - 4$.
Taking into account the extra minus sign,
$\Omega'$ now acts as $e^{i\pi (q_R/2+1)}\,=\,e^{i\pi( \ell+1)}$,
leaving only states with $q_R=2$ mod 4 or $\ell=1$ mod 2:
\be
&&\Delta= 3\qquad\qquad (1,3)_2, \quad m^2\,=\,-3,\nn\\
&&\Delta= 5\qquad\qquad (1,3)_6, \quad m^2\,=\,5,\nn\\
&&\qquad\,\,\vdots
\ee
The scalars $(1,1;1,1)$ gave rise to states
$(1,1)_{2\ell}$, $\ell \geq 2$,
with $m^2=\ell^2-4\ell$,
which satisfy the constraint \eqs{chiralprimaries}.
$\Omega'$ acts on these states as $e^{i\pi q_R/ 2}\,=\,e^{i\pi \ell}$, 
which includes the extra minus sign coming from $\Omega(-1)^{F_L}$. 
Only states with  $q_R=0$ mod 4, or $\ell=0$ mod 2, survive:
\be
&&\Delta=2\qquad (1,1)_4,\quad m^2=-4, \nn\\
&&\qquad\,4\qquad (1,1)_8,\quad m^2=0,\ret
&&\qquad\,\,\vdots
\ee
We include these states in Table {\tblspxsp} below. 

\noindent (2) {\it Vector modes:}

\noindent The reduction on $S^1$ of the six-dimensional tensor field
yields a massless vector state with 
$\su{2}_L\times\su{2}_R\times\u{1}_R$ quantum numbers $(1,1)_0$.
This state, however, is projected out by $\Omega'$,
consistent with the fact that 
there is no global symmetry current 
other than the $\su{2}_L\times\su{2}_R\times\u{1}_R$ currents 
from the bulk and the $\so{4}\times\so{4}$ currents from the D7-brane sector.

\vs{.1in}
\noindent {\large\bf (c) D7-brane sector:}
\vs{.1in}

\noindent The D7-brane sector consists of the world-volume theory of 8 D7-branes
with topology
$AdS_5\times S^3$, modded out by
$\IZ_2^{\rm orb}$. 
The $\IZ_2^{\rm orb}$ projection differs from
that in the $\su{N}+2\,\Yasymm+4\,\fund$ case, 
the corresponding projection matrix being -- see Eq. \eqs{project2}:
\be
\gamma_{\theta 7}^{\rm Sp\times Sp}= \Biggr(\begin{tabular}{cc}
\hskip-2pt $\identity_4$ &
\hskip-8pt 0
\\[-.05in]
\hskip-7pt 0 & \hskip-10pt $-\identity_4$
\end{tabular}
\hskip-5pt\Biggr),
\la{projectspsp}
\ee
which breaks the $\so{8}$ gauge group on
the D7-branes to $\so{4}\times\so{4}$. 
Accordingly, the adjoint of $\so{8}$
decomposes as 
$\rep{28}\to(6,1)\oplus(1,6)\oplus (4,4)$,
where $(6,1)\oplus(1,6)$ is the adjoint representation of $\so{4}\times\so{4}$.
In analogy with the $\su{N}+2\Yasymm$ case, 
D7-brane excitations that are in odd
representations of $\su{2}_L$ transform 
in the $(6,1)\oplus(1,6)$ of the flavor group, 
while excitations in even representations of $\su{2}_L$
transform in the
$(4,4)$ of the flavor group. 
These results are summarized in Table {\tblspxsp}. 

\vbox{
\begin{center}
\begin{tabular}{||c|c|c|c|c|c|c||}
\hline\hline
$\Delta$ & Sector & $\su{4}_R$ & 
$(d_L, d_R)_{q_R}$ & 
$\so{4}\times\so{4}$ & $m^2$ &  CFT operator  
\\
\hline\hline
2 & bulk& $\rep{20'}$ & $(1,1)_4$ &\rule{5mm}{.1mm}& $-4$ &
	  tr\,$\Phi_1{}^2\,+\, {\rm tr}\,\Phi_2{}^2$ \\  
2 & bulk& $\rep{20'}$ & $(3,3)_0$ &\rule{5mm}{.1mm}& $-4$ & 
	  tr\,$[B_{(i}J_1B_{j)}J_2]$ \\ 
2 & twisted & & $(1,1)_4$ &\rule{5mm}{.1mm}& $-4$ &  
	${\rm tr} \,\Phi_1{}^2\,-\, {\rm tr}\,\Phi_2{}^2$ \\ 
2 & D7 &  &  $(1,3)_0$ & $(6,1)\oplus(1,6)$ & $-4$ & 
	$Q_1^{[I}J_1Q_1^{K]}\oplus Q_2^{[I}J_2Q_2^{K]}$ \\ 
3 & bulk& \rep{10} & $(1,3)_2{}$ &\rule{5mm}{.1mm}& $-3$ &  
	$ {\rm tr}\,\lambda_1{}^2$ + $ {\rm tr}\,\lambda_2{}^2$ \\ 
3 & twisted & & $(1,3)_2{}$ &\rule{5mm}{.1mm}& $-3$ &  
	$ {\rm tr}\,\lambda_1{}^2-  {\rm tr}\,\lambda_2{}^2$   \\ 
3 & D7 &  & $(2,4)_0$ & $(4,4)$ & $-3$ & $Q^{I}_1J_1 B_iJ_2 Q^{K}_2 $
\\  
4 & bulk& \rep{105} & $(1,1)_8$ &\rule{5mm}{.1mm}& 0 &
      ${\rm tr} \,\Phi_1{}^4\,+\, {\rm tr}\,\Phi_2{}^4$ \\ 
4 & bulk& \rep{105} & $(3,3)_4$ &\rule{5mm}{.1mm}& 0 & 
   tr  $[B_{(i}J_1\Phi_1{}^2J_2B_{j)}]   \pm 
   {\rm tr} [B_{(i}J_1\Phi_2{}^2J_2B_{j)}]$ \\
4 & bulk& \rep{105} & $(5,5)_0$ &\rule{5mm}{.1mm}& 0 & 
	tr  $[J_1 B_{(i} J_2 B_jJ_1 B_k J_2 B_{l)}]$ \\ 
4 & twisted & & $(1,1)_8$ &\rule{5mm}{.1mm}& 0 &  
	${\rm tr} \,\Phi_1{}^4\,-\, {\rm tr}\,\Phi_2{}^4$ \\ 
4 & D7 &  &  $(3,5)_0$ & $(6,1)\oplus(1,6)$ & 0 &  
	$Q^{[I}_1 J_1 B_{(i} J_2 B_{j)} J_1 Q^{K]}_1
	\oplus  Q^{[I}_2 J_2 B_{(i} J_1 B_{j)} J_2 Q^{K]}_2 $ \\
\hline\hline
\end{tabular}
\end{center}
\begin{center}
{\small{\bf Table {\tblspxsp}}: 
Scalars  with $\Delta \le 4$ in chiral primary representations of \\
$\sp{2N}_1\times\sp{2N}_2$\,w/\,
$(\fund{}_{1},\fund{}_{2})\oplus2\,\fund\,{}_{1}\oplus 2\,\fund\,{}_{2}$
\cite{GukovKapustin}.}
\end{center}
}
\vspace{0.5cm}

The last column of Table {\tblspxsp} 
lists the CFT operators that correspond 
to the supergravity states.
The operators are written as products of fields in the
perturbative $\sp{2N}\times \sp{2N}$ worldvolume theory,
listed in Table {\tblspxspfield}.
(It is difficult to specify the CFT operators unambiguously
without explicitly writing all the gauge theory indices;
instead, we adopt a slightly schematic notation,
in which $J_1$ and $J_2$ 
denote the invariant antisymmetric tensor of $\sp{2N}$,
acting on the indices of the first and second factor 
of the gauge group respectively.)

\vbox{
\begin{center}
\begin{tabular}{||l|c|c|c|c||}
\hline\hline
$\Delta$ & $\sp{2N}\times \sp{2N} $ 
& $\su{2}_L \times \su{2}_R \times \u{1}_R $ & 
$\so{4}\times\so{4}$ & CFT field  
\\
\hline\hline
${3\over 2}$ 	& (adjoint,1) 	& $(1,2)_1$ & (1,1) & $\lambda_1$ \\ 
${3\over 2}$ 	& (1, adjoint) 	& $(1,2)_1$ & (1,1) & $\lambda_2$ \\  
{1} 		& (adjoint,1) 	& $(1,1)_2$ & (1,1) & $\Phi_1$ \\  
{1} 		& (1, adjoint) 	& $(1,1)_2$ & (1,1) & $\Phi_2$ \\  
{1} 		&$(\fund,\fund)$& $(2,2)_0$ & (1,1) & $B_i$ \\  
{1} 		&$ (\fund, 1)$ 	& $(1,2)_0$ & (4,1) & $Q_1^I $ \\  
{1} 		& $(1, \fund)$ 	& $(1,2)_0$ & (1,4) & $Q_2^I $ \\  
\hline\hline
\end{tabular}
\end{center}
\begin{center}
\baselineskip12pt
{\noindent\small{\bf Table {\tblspxspfield} }: 
$\su{2}_R$ multiplets in perturbative
$\sp{2N}_1\times\sp{2N}_2$\,\, w/\,\,
$(\,\fund\,{}_{1},\fund\,{}_{2})\,\oplus\,2\,\fund\,{}_{1}\oplus
\,2\,\fund\,{}_{ 2}$
}
\end{center}
}
\vspace{0.5cm}

The $\su{2}_L$ factor of the global symmetry
corresponds to the flavor symmetry of the bifundamental hypermultiplet.
The explanation is similar to that 
in the $\sp{2N}  + \Yasymm + 4\, \fund$ case. 
The bifundamental hypermultiplet contains  chiral superfields 
$B_{a_1}{}^{a_2}$ in the $(\fund_1,\overline{\fund}_2)$ and
$\tilde B^{a_1}{}_{a_2}$ in the $(\overline{\fund}_1,\fund_2)$ of
$\sp{2N}_1\times\sp{2N}_2$.
The piece of the $\C{N}=2$ superpotential involving the first $\sp{2N}$ 
factor, say, is 
\be
 \tilde B^{a_1}{}_{a_2}\Phi_{a_1}{}^{b_1} B_{b_1}{}^{a_2},
\la{supermann}
\ee
where $\Phi_{a_1}{}^{b_1}$ is a chiral superfield in the adjoint of
$\sp{2N}_1$. We can define two chiral superfields 
$B_{1\,a_1a_2}\equiv J_{a_2b_2}B_{a_1}{}^{b_2}$ and
$B_{2\,a_1a_2}\equiv J_{a_1 b_1}\tilde B^{b_1}{}_{a_2}$,
both transforming as $(\fund_1,\fund_2)$. 
In terms of these new fields the superpotential reads
\be
J^{a_2 b_2}B_{2\,a_1a_2}\widehat\Phi^{a_1 b_1} B_{1\,b_1b_2},
\la{supermann2}
\ee
where $\widehat\Phi^{a_1 b_1}=J^{c_1 a_1}\Phi_{c_1}{}^{b_1}$ is
symmetric in $a_1, b_1$. 
Since eq.~\eqs{supermann2} is odd under exchange 
$1\leftrightarrow 2$, 
the superpotential actually possesses an $\su{2}_L$ flavor symmetry
acting on the doublet $B_i=(B_1, B_2)$. 
A bare mass $\globalmass$ ($\equiv$ global mass)
for the bifundamental  hypermultiplet $\globalmass B^2$ 
breaks ${\rm SU}(2)_L\rightarrow \u{1}_L$.

\subsection{$\su{N}$ + 1 antisymmetric + 1  symmetric hypermultiplet}

We now move on to the models coming from the Type IIA
O$6^{+}$--\,O$6^{-}$ configuration.
As in the O$6^{-}$--\,O$6^{-}$ case, 
we obtain two theories with almost the same IIB description,
namely $\su{N}$ with symmetric and antisymmetric hypermultiplets, 
and $\sp{2N}\times\so{2N+2}$ with a bifundamental hypermultiplet. 
In the near-horizon supergravity description, 
both theories share the same bulk sector,
and are only distinguished by the twisted sector of the
$\IZ_2^{\rm orb}$ orbifold T-dual to the NS 5-branes. 
{}From Sect.~2.2, we recall that the orientifold group is 
\be
G_{\rm orient} = \IZ_{2}^{\rm orb} + \IZ_{2}^{\rm orb}\alpha\,
\Omega',
\la{orientiplusplus2}
\ee
where, as in previous cases, 
$\Omega'=R_{45}\,\Omega\,(-1)^{F_L}$, 
and 
$\IZ_2^{\rm orb } = \{\identity, \alpha^2\}$, 
with 
$\alpha^2=R_{6789}$. 
$G_{\rm orient}$ is actually a $\IZ_4$ group with
elements $\{\identity, \alpha\Omega', \alpha^2, \alpha^3\Omega'\}$.
Although \eqs{orientiplusplus2} is superficially analogous to the
orientifold group of the O$6^{-}$--\,O$6^{-}$ models, the relevant
fixed-point set of the $G_{\rm orient}$ action reduces in this case to the
six-dimensional hyperplane
$x^6=x^7=x^8=x^9=0$ left fixed by
$\IZ_2^{\rm orb}$, as the orientifold factor $\alpha\Omega'$ only fixes the
origin of the space transverse to the D3-branes,
$x^4=x^5=x^6=x^7=x^8=x^9=0$, 
which does not give any contribution in the near-horizon limit. 
This is as expected, however, 
as we know that the T-dual of the O$6^{+}$--\,O$6^{-}$ configuration 
contains no D7-branes, 
which would be present had there been non-trivial fixed points
associated with the $\Omega'$ action. 

In the near-horizon limit the space-time geometry is $AdS_5\times
S^5/G_{\rm orient}$, with the following metric for the compact space  
\be
&&d\widetilde{
\Omega}_5{}^2 = d\theta_1{}^2 + 
\cos^2\theta_1 \,\left(d\theta_2{}^2 +
\cos^2\theta_2\,d\phi_1{}^2+\sin^2\theta_2\,d\phi_2{}^2\,\right)+
\sin^2\theta_1\,
d\phi_3{}^2,\ret &&\theta_{1,2}\sim 
\theta_{1,2}+\frac{\pi}{2},\quad
\phi_{1}\sim \phi_{1}+\frac{\pi}{2}, \quad \phi_{2}\sim
\phi_{2}-\frac{\pi}{2},\quad \phi_{3}\sim \phi_{3}+\pi.  
\la{metricanti+symm}
\ee

The isometry group of the above metric is a quotient of 
$\u{1}_L \times\su{2}_R\times\u{1}_R$,
where $\u{1}_L\subset\su{2}_L$, 
which we can define as follows.
$\alpha^2$ acts on $\su{2}_L$ representations as $e^{i\pi J_L}$, with
$J_L/2$ the spin of the representation ($J_L=1$ for the $\rep2$, $J_L=2$
for the $\rep3$, and so on), so it projects out even-dimensional
representations, thereby breaking
$\su{2}_L\to\so{3}_L$. $\alpha$ acts on $\so{3}_L$ representations 
according to the weight of each state (or $\u{1}_L\subset\so{3}_L$ charge
$q_L$) as $e^{i\pi q_L/2}$, where $q_L=\pm2,0$ for the $\rep3$, $\pm4,
\pm2,0$ for the $\rep5$, and so on, so that on $\so{3}_L$ representations
$\alpha$ acts as $\pm 1$. This immediately suggests  an alternative way to
regard
$G_{\rm orient}$ 
that will prove useful in understanding the twisted sector of the spectrum, 
namely, to consider $G_{\rm orient}$ as 
$\IZ_2^{\rm orb} \times (G_{\rm orient}/\IZ_2^{\rm orb})$
where the coset
$G_{\rm orient}/ \IZ_2^{\rm orb} \sim \{\identity, \alpha\Omega'\}$. 
In other words, we mod out in two stages, first by $\IZ_2^{\rm orb}$, 
and then by $\alpha\Omega'$.
Both approaches give the same result for the bulk sector, 
while the second one makes the analysis of the twisted sector straightforward.  Finally,
$R_{45}$ acts on states with
$\u{1}_R$ charge $q_R$ as $e^{i\pi q_R/2}$. 
It is important to keep in mind when  performing 
the projection that
$G_{\rm orient}$ has a single generator $\alpha\Omega'$ 
(and will therefore constrain the allowed values of $q_L+q_R$). 
This differs from the projection in the
$\su{N}+2\,\Yasymm+4\,\fund$ or $\sp{2N}\times\sp{2N}$  cases, 
in which the orbifold and orientifold factors act independently, 
so that the $\su{2}_L$ representation and $\u{1}_R$ charges 
of the states that survive the projection obey independent constraints. 

The supergravity spectrum includes states from two sectors:
(a) the bulk sector, consisting of $AdS_5\times S^5$ supergravity states 
invariant under $G_{\rm orient}$,
and 
(b) the $\IZ_2^{\rm orb}$ twisted sector, 
containing states twisted with respect to $\alpha^2$, 
and modded further by $\alpha\Omega'$.

\vs{.1in}
\noindent{\large\bf (a) Bulk sector:}
\vs{.1in}

\noindent The strategy to follow is essentially the same as in the
$\su{N}+2\,\Yasymm+4\,\fund$ case, 
but with a few differences that we will illustrate with two examples. 

\noindent (1) {\it Scalar modes: }

\noindent Consider first the supergravity mode with $m^2 = -4$
in the ${\bf 20'}$ of $\su{4}_R$, 
which couples to a $\Delta=2$ operator on the boundary. 
Under the  branching  ${\rm SO}(6)
\rightarrow {\rm SU}(2)_L \times\su{2}_R \times \u{1}_R$, this
representation decomposes as (see Table
{\tblbranching})
\be
\rep{20'} \rightarrow
(3,3)_0+(2,2)_2+(2,2)_{-2}+(1,1)_4+(1,1)_{-4}+(1,1)_0.
\ee
$\IZ_2^{\rm orb}$ projects out all the states in even-dimensional
representations of $\su{2}_L$, 
leaving $(3,3)_0+(1,1)_4+(1,1)_{-4}+(1,1)_0$. 
We now split the $\su{2}_L$ multiplets according to 
their $\u{1}_L$ charge $q_L$, so that, for example, 
$(3,3)_0\to3_0{}^{\pm2}+3_0{}^0$, where the superscript denotes $q_L$.
We now keep only those states invariant under $\alpha\Omega'$. 
Since $\alpha$ acts as $e^{i\pi q_L/2}$ 
and the geometric part of $\Omega'$ as $e^{i\pi q_R/2}$, 
the surviving states must satisfy $q_L +q_R=0\mod4$. This leaves
\be
3_0{}^0,\quad 1_{\pm4}{}^0,\quad 1_{0}{}^0,
\ee
of which only $3_0{}^0$ and $1_{4}{}^0$ obey the constaint \eqs{chiralprimaries}
and so correspond to chiral primary states of the CFT. 
($1_{-4}{}^0$ is the complex conjugate of $1_{4}{}^0$, and is
an anti-chiral primary state.)

There are two modes with $m^2 = -3$, 
in the ${\bf 10}$ and the ${\bf 50}$ of $\su{4}_R$, 
which couple to $\Delta=3$ operators on the boundary.
The ${\bf 50}$ decomposes into
\be
\rep{50} \rightarrow
(4,4)_0+(3,3)_{\pm 2}+(2,2)_{\pm 4}+(2,2)_0+(1,1)_{\pm 6}+(1,1)_{\pm 2}
\ee
Projecting by $\alpha^2$ removes the even-dimensional 
representations of $\su{2}_L$, leaving
$(3,3)_{\pm 2}+(1,1)_{\pm 6}+(1,1)_{\pm 2}$. 
Breaking the resulting $\so{3}_L$ to $\u{1}_L$,
and projecting by $\alpha\Omega'$ leaves states with $q_L +q_R=0\mod4$, 
namely,
\be
3_2{}^{\pm2},\quad 3_{-2}{}^{\pm2},
\ee
of which $3_2{}^{\pm2}$ are chiral primaries 
(while $3_{-2}{}^{\pm2}$ are 
the complex conjugate anti-chiral primaries).
The ${\bf 10}$ decomposes into 
\be
\rep{10} \rightarrow (1,3)_2 +(3,1)_{-2} +(2,2)_0
\ee
$\alpha^2$ removes $(2,2)_0$, while $\alpha\Omega'$ keeps only
those states with $q_L+q_R=2\mod 4$, 
taking into account the extra minus sign 
from the action of $\Omega(-1)^{F_L}$ on the ten-dimensional fields. 
This leaves $3_2{}^0$ and $1_{-2}{}^0$, 
of which only $3_2{}^0$ is a chiral primary.

Proceeding in this way, one finds the results summarized in Table
{\tblsuantisym}. 

\noindent (2) {\it Vector modes: }
 
\noindent  Start with the  
massless vector mode in the {\bf 15} of ${\rm SO}(6)$ 
(see Table \tablevectors), which couples to the 
$\Delta=3$ $\su{4}_R$ current on the boundary.  
Given the  branching 
\be
\rep{15} \rightarrow (1,1)_0+(2,2)_2+(2,2)_{-2}+(3,1)_0+(1,3)_0, 
\ee
we see that  $\alpha^2$ projects out the $(2,2)_{\pm2}$, 
while the $\alpha\Omega'$ projection leaves
three massless vector states
$1_0{}^0,~3_0{}^0,~1_0{}^0$, 
which couple to the
$\u{1}_L\times\su{2}_R\times\u{1}_R$ currents of the boundary CFT.

\vs{.1in}
\noindent {\large\bf (b) $\IZ_2^{\rm orb}$ twisted sector:}
\vs{.1in}

\noindent 
As in the $\su{N}+2\,\Yasymm+4\,\fund$ or $\sp{2N}\times\sp{2N}$  cases,
we start with
the Type IIB theory on $\IC^{2}_{6789}/\IZ_{2}^{\rm orb}$.
The twisted sector consists of a $(2,0)$ $d=6$ tensor multiplet, 
which we have to mod out by $\alpha \Omega'$.
The tensor multiplet is a singlet under $\su{2}_L$ (see Table {\tbltensor}), 
and therefore insensitive to the action of $\alpha$, 
so we only have to project by $\Omega'$, 
exactly as in the $\su{N}+2\,\Yasymm+4\,\fund$ or $\sp{2N}\times\sp{2N}$ cases. 
As in those cases, there are two different ways to
perform the projection \cite{ParkUranga}. 
The projection giving the $\sp{2N}\times \so{2N+2}$ twisted sector 
(which is the same as that for $\sp{2N}  \times \sp{2N}$ )
has an extra minus sign relative to that giving the 
$\su{N}+\Yasymm+\Ysymm$ twisted sector 
(which is the same as that of $\su{N}+2\,\Yasymm+4\,\fund$). 
We have come to the conclusion that the contribution to the supergravity
spectrum of the $\IZ_2^{\rm orb}\,$ twisted sector 
for the $\su{N}+\Yasymm+\Ysymm$ theory
is exactly the same as that of the $\su{N}+2\,\Yasymm+4\,\fund$ theory, 
which we derived in section \ref{sect2anti}. 
(The interpretation in terms of field theory operators is of course 
different.)

\vbox{
\begin{center}
\begin{tabular}{||c|c|c|c|c|c||}
\hline\hline
$\Delta$ & Sector & $\su{4}_R$ & 
$(d_R)_{\,q_R}{}^{\!\!q_L}$ &
$m^2$ & CFT operator \\
\hline\hline
2 & bulk & $\rep{20'}$ & $1_4{}^0\,$ & $-4$ & ${\rm tr} \,\Phi^2$ 
\\  
2 & bulk & $\rep{20'}$ & $\,3_0{}^0$ & $-4$ & $\tilde A A-\tilde S S$ 
\\ 
2 & twisted & & $\,3_0{}^0$ & $-4$ & $\tilde A A + \tilde S S$ 
\\ 
3 & bulk & $\rep{10}$ & $3_2{}^0$ & $-3$ & $ {\rm
tr}\,\lambda^2$ 
\\ 
3 & bulk & \rep{50} & $3_2{}^{ 2}$ & $-3$ & $\tilde S \Phi A$ 
\\ 
3 & bulk & \rep{50} & $3_2^{-2}$ & $-3$ & $\tilde A \Phi S$  
\\  
3 & twisted & & $1_6{}^0$ &  $-3$ & ${\rm tr}\, \Phi^3$ 
\\ 
 & bulk & \rep{105} & $5_0{}^{4}$ & 0 & $(\tilde S A)^2$ 
\\
4 & bulk & \rep{105} & $5_0^{- 4}$ & 0 & $(\tilde  A S)^2$ 
\\ 
4 & bulk & \rep{105} & $5_0{}^{0}$ & 0 & $(\tilde  A A\pm\tilde S S)^2$ 
\\ 
4 & bulk & \rep{105} & $\,3_4{}^{0}$ & 0 & $\tilde A \Phi^2 A-\tilde S \Phi^2 S$
\\ 
4 & bulk & \rep{105} & $1_8{}^0$ & 0 & ${\rm tr}\, \Phi^4$ 
\\ 
4 & twisted & & $\,3_4{}^0$ & 0 & $\tilde A \Phi^2 A+ \tilde S
\Phi^2 S$ 
\\ 
\hline\hline
\end{tabular}
\end{center}
\begin{center}
\begin{minipage} {5.5in}
\baselineskip12pt
{\noindent\small{\bf Table {\tblsuantisym}}: 
Scalars with $\Delta \le 4$ in chiral primary representations of 
$\su{N}$\,\, w/\,\,\,$\Ysymm\oplus\Yasymm\,$.  The dimension of
of the $\su{2}_R$ representation is denoted by $d_R$, while
$q_R$, $q_L$ denote the $\u{1}_R$, $\u{1}_L$ charges.}
\end{minipage}
\end{center}
}

The results are summarized in Table {\tblsuantisym}. 
The last column of Table {\tblsuantisym} 
lists the CFT operators that we conjecture 
correspond to the supergravity states (see the discussion below).
The operators are written as products of fields in the
perturbative $\su{N}+\Ysymm+\Yasymm$ worldvolume theory,
listed in Table {\tblsuantisymfield}.

\vbox{
\begin{center}
\begin{tabular}{||l|c|c|c|c||}
\hline\hline
$\Delta$ & $\su{N}$ & $\su{2}_R \times \u{1}_R \times \u{1}_L $ & 
$\widetilde{\u{1}}_L$  &
 CFT field  
\\
\hline\hline
${3\over 2}$ 	&  adjoint 	& $2_1{}^0$  	& 0 & $\lambda$ \\  
{1} 		& adjoint 	& $1_2{}^0$  	& 0 & $\Phi$ \\  
{1}		& $\Yasymm$ 	& $2_0{}^1$  	& 1 & $A$ \\  
{1} 		& $\Ysymm$ 	& $2_0^{-1}$   	& 1 & $S$ \\  
\hline\hline
\end{tabular}
\end{center}
\begin{center}
\baselineskip12pt
{\noindent\small{\bf Table {\tblsuantisymfield} }: 
$\su{2}_R$ multiplets in perturbative
$\su{N}$\,\, w/\,\,\,$\Ysymm\oplus\Yasymm\,$
}
\end{center}
}

We recall that among the states 
from the $\IZ_2^{\rm orb}$ twisted sector
is a massless vector with quantum numbers $1_0{}^0$ 
that couples to a conserved current on the boundary. 
Let us denote the corresponding global (abelian) symmetry group 
by $\widetilde{\u{1}}_L$. 
We now consider how this $\widetilde{\u{1}}_L$,
together with the $\u{1}$ symmetry from the bulk sector,
acts on the fields of the theory.  

The $\su{N}+\Yasymm+\Ysymm$ theory contains hypermultiplets $\{A,
\tilde A\}\,,
\{S,
\tilde S\}$ in the antisymmetric and symmetric representations of $\su{N}$.
The superpotential is
\be
\C{W}={\rm tr}\,(\,\tilde A
\Phi A\,+\,\tilde S
\Phi S\,),
\ee which possesses a $\u{1}_A\times \u{1}_S$ flavor symmetry acting
as follows: 

\begin{equation}
\begin{tabular}{lllll}
\multicolumn{2}{c}{$\u{1}_A$} &{}& \multicolumn{2}{c}{$\u{1}_S$} \\
$A \rightarrow e^{i\alpha} A$ & $\tilde A \rightarrow e^{-i\alpha}
\tilde A$ & {} &
$A \rightarrow  A$ & $\tilde A \rightarrow 
\tilde A$
\\
$S \rightarrow S$ & $\tilde S \rightarrow 
\tilde S $ & {} &
$S \rightarrow e^{i\beta} S$ & $\tilde S \rightarrow
e^{-i\beta} \tilde S$
\end{tabular}
\la{uunosantisymm1}
\end{equation}
\vspace{.1in}

\noindent 
We would like to identify $\u{1}_L$ and  $\widetilde{\u{1}}_L$ 
in terms of $\u{1}_A$ and $\u{1}_S$. 
We can give masses 
$m_A\tilde A A\,+\,  m_S\tilde S S$ to both hypermultiplets 
without breaking $\u{1}_A \times \u{1}_S$. 
Defining $\globalmass = (m_A-m_S)/2$ (global mass)
and $\mu=(m_A+m_S)/2$,  we rewrite the mass term as 
\be
\globalmass\,(\tilde A A-\tilde S S)\,+\,\mu\,(\tilde A A +\tilde S S).
\ee
Following the discussion of the $\su{N}+ 2\,\Yasymm + 4\,\fund$ theory,
it seems reasonable to identify the $\u{1}_L$ and $\widetilde{\u{1}}_L$ 
generators, $T_L$ and $\widetilde T_L$, as
\be
T_L\,=\,T_A-T_S ,\qquad\widetilde T_L\,=\,\tilde T_A+\tilde T_S
\ee
where $T_A$ and $T_S$ are the generators of $\u{1}_A$ and $\u{1}_S$.
Thus, $\u{1}_L\times\widetilde{\u{1}}_L$ acts  on the hypermultiplets as

\begin{equation}
\begin{tabular}{lllll}
\multicolumn{2}{c}{$\u{1}_L$} &{}&
\multicolumn{2}{c}{$\widetilde{\u{1}}_L$} \\
$A \rightarrow e^{i\alpha} A$ & $\tilde A \rightarrow e^{-i\alpha}
\tilde A $ & {} &
$A\rightarrow e^{i\beta} A$ & $\tilde A\rightarrow e^{-i\beta}
\tilde A$
\\
$S\rightarrow e^{-i\alpha} S$ & $\tilde S\rightarrow
e^{i\alpha}
\tilde S$ & {} &
$S \rightarrow e^{i\beta} S$ & $\tilde S \rightarrow e^{-i\beta}
\tilde S$
\end{tabular}
\la{uunosantisymm2}
\end{equation}
\noindent which we include in Table {\tblsuantisymfield}.

Further, based on analogy with the $\su{N}+ 2\,\Yasymm + 4\,\fund$ theory,
it seems reasonable to identify the operator corresponding to
deformation of the global mass, $\tilde A A-\tilde S S$,
with the $3_0{}^0$ supergravity operator in the bulk sector,
and the operator corresponding to deformation of the average mass $\mu$
of the hypermultiplets, $\tilde A A+\tilde S S$,
with the $3_0{}^0$ supergravity operator in the twisted sector.
This assignment, however, can only be regarded as conjectural.
(Similar remarks hold for 
some of the other entries of Table {\tblsuantisym}.)

\subsection{$\sp{2N}\times\so{2N+2}$ +  1 bifundamental hypermultiplet }

The analysis of the 
$\C{N}=2$ $\sp{2N}\times\so{2N+2}$  theory with one
bifundamental hypermultiplet
proceeds in parallel with the previous one, 
as they both share the same IIB background 
(with a few differences that we have already discussed
which serve to distinguish the two). 

The orientifold group and metric are the same as in the previous case
(see Eqs.~\eqs{orientiplusplus2}  and \eqs{metricanti+symm}). 
As before, there is a six-dimensional
hyperplane left fixed by the $G_{\rm orient}$,
which becomes $AdS_5\times S^1$ in the near-horizon limit. 

The supergravity spectrum consists of two sectors: 

\vs{.1in}
\noindent{\large\bf (a) Bulk sector:}
\vs{.1in}

\noindent This sector consists of $AdS_5\times S^5$ supergravity states
invariant under the orientifold group \eqs{orientiplusplus2},
and is identical to that in the $\su{N}+\Ysymm+\Yasymm$ theory, 
although the field theory interpretation of the 
resulting states obviously differs (see Table {\tblspxso}).

\vs{.1in}
\noindent{\large\bf (b) $\IZ_2^{\rm orb}$ twisted sector:}
\vs{.1in}

\noindent The $\IZ_2^{\rm orb}$ twisted sector, 
supported on $AdS_5\times S^1$, 
consists of the KK reduction on $S^1$ 
of the six-dimensional $(2,0)$ tensor multiplet,
modded out by $\alpha\Omega'$.

\vbox{
\begin{center}
\begin{tabular}{||c|c|c|c|c|c||}
\hline\hline
$\Delta$ & Sector & $\su{4}_R$ & 
$(d_R)_{\,q_R}{}^{\!\!q_L}$ 
& $m^2$ & CFT operator  
\\
\hline\hline
2 & bulk & $\rep{20'}$ & $1_4{}^0\,$ & $-4$ & ${\rm tr} \,\Phi^2_{\rm Sp}\,\pm\,
{\rm tr}\,\Phi^2_{\rm SO}$
\\  
2 & bulk & $\rep{20'}$ & $\,3_0{}^0$ & $-4$ & $\tilde B B $ 
\\ 
2 & twisted & & $\,1_4{}^0$ & $-4$ & ${\rm tr}\,\Phi^2_{\rm
Sp}\,\mp\,{\rm tr} \,\Phi^2_{\rm SO}$ 
\\ 
3 & bulk & \rep{10} & $3_2{}^0$ & $-3$ & ${\rm tr} \,\lambda^2_{\rm Sp}
\,\pm\, {\rm tr}\,\lambda^2_{\rm SO}$ 
\\ 
3 & bulk & \rep{50} & $3_2{}^{ 2}$ & $-3$ & $B(J\Phi_{\rm Sp})B\pm 
B \Phi_{\rm SO}(JB)$ 
\\
3 & bulk & \rep{50} & $3_2^{- 2}$ & $-3$ & 
  $\tilde B (J \Phi_{\rm Sp}) \tilde B \pm \tilde B \Phi_{\rm SO} (J\tilde B)$ 
\\ 
3 & twisted & & $3_2{}^0$ & $-3$ & ${\rm tr} \,\lambda^2_{\rm Sp}
\,\mp\, {\rm tr}\,\lambda^2_{\rm SO}$ 
\\ 
4 & bulk & \rep{105} & $5_0{}^{ 4}$ & 0 & $BJBBJB$ 
\\
4 & bulk & \rep{105} & $5_0^{-4}$ & 0 & $\tilde BJ\tilde B\tilde BJ\tilde B$ 
\\ 
4 & bulk & \rep{105} & $5_0{}^{0}$ & 0 & $(\tilde  B B)^2$ 
\\ 
4 & bulk & \rep{105} & $\,3_4{}^{0}$ & 0 & $\tilde  B \Phi^2_{\rm Sp} B \pm 
\tilde B \Phi^2_{\rm SO} B$ 
\\ 
4 & bulk & \rep{105} & $1_8{}^0$ & 0 & ${\rm tr}\, \Phi^4_{\rm Sp}\,\pm\,
{\rm tr}\,\Phi^4_{\rm SO}$ 
\\ 
4 & twisted & & $\,1_8{}^0$ & 0 & ${\rm tr}\,
\Phi^4_{\rm Sp}\,\mp\,{\rm tr}\,\Phi^4_{\rm SO}$ 
\\
\hline\hline
\end{tabular}
\end{center}
\begin{center}
\baselineskip12pt
{\noindent\small{\bf Table {\tblspxso}}: 
Scalars with $\Delta \le 4$ in chiral primary representations of
$\sp{2N} \times \so{2N+2}$\,\, w/\,\,
$(\,\fund\,,\fund\,)$ }
\end{center}
}

As $\alpha$ acts trivially on the
tensor multiplet, which is an $\su{2}_L$ singlet,
we need only project by $\Omega'$. 
This sector is therefore identical to that 
in the $\sp{2N}\times\sp{2N}$ theory (at least for large $N$). 
Recall that $\Omega'$ acts on the twisted sector 
with an additional minus sign relative to the 
$\su{N}+2\,\Yasymm+4\,\fund$ or $\su{N}+\Ysymm+\Yasymm$ theories.

The spectrum of scalars with $\Delta \le 4$
in chiral primary representations
is summarized in Table {\tblspxso}.
The last column of Table {\tblspxso}
lists the CFT operators that correspond 
to the supergravity states.
The operators are written as products of fields in the
perturbative $\sp{2N}\times \so{2N+2}$ worldvolume theory,
listed in Table {\tblspxsofield}.
As in the previous section, 
in many cases there are several different CFT operators 
with the correct quantum numbers to match the supergravity states, 
and we do not even attempt to resolve the ambiguities.

\vbox{
\begin{center}
\begin{tabular}{||l|c|c|c||}
\hline\hline
$\Delta$ & $\sp{2N}\times\so{2N+2} $ 
& $\su{2}_R \times \u{1}_R  \times \u{1}_L$ & CFT field  
\\
\hline\hline
${3\over 2}$ 	& (adjoint,1) 	& $2_1{}^0$ & $\lambda_{\rm Sp}$ \\ 
${3\over 2}$ 	& (1, adjoint) 	& $2_1{}^0$ & $\lambda_{\rm SO}$ \\  
{1} 		& (adjoint,1) 	& $1_2{}^0$ & $\Phi_{\rm Sp}$ \\  
{1} 		& (1, adjoint) 	& $1_2{}^0$ & $\Phi_{\rm SO}$ \\  
{1} 		&$(\fund,\fund)$& $2_0{}^1$ & $B$ \\  
\hline\hline
\end{tabular}
\end{center}
\begin{center}
\baselineskip12pt
{\noindent\small{\bf Table {\tblspxsofield} }: 
$\su{2}_R$ multiplets in perturbative
$\sp{2N} \times \so{2N+2}$\,\, w/\,\, $(\,\fund\,,\fund\,)$. 
}
\end{center}
}
\vspace{0.5cm}

\begin{center}
{\bf ~Acknowledgements} 
\end{center}
We would like to thank E. Cortegoso for assistance with the figures.

\end{document}